\def\Journal#1#2#3#4{{#1} {\bf #2}, #3 (#4)}

\def\NPB{Nucl. Phys. }
\def\PLB{Phys. Lett. }
\def\PRL{Phys. Rev. Lett.}
\def\PRD{Phys. Rev. D}
\def\ZPC{{Z. Phys.} C}

\documentstyle[aps,manuscript,epsf,rotate]{revtex}
\begin{document}

\title{
	\rightline{\small KANAZAWA 99-07, HUB-EP 99/19} 
       Monopole characteristics in various Abelian gauges}
 
\author{
    E.-M. Ilgenfritz \\
    {\it Institute for Theoretical Physics, Kanzawa University, Japan} \\
       S. Thurner, H. Markum \\
    {\it Institut f\"ur Kernphysik, TU-Wien, Austria }\\
       M. M\"uller--Preussker \\
    {\it  Institut f\"ur Physik, Humboldt--Universit\"at zu Berlin, Germany}
}
\maketitle

\begin{abstract}
Renormalization group (RG) smoothing is employed on the lattice 
to investigate and
to compare the monopole structure of the $SU(2)$ vacuum as seen in different 
gauges (maximally Abelian (MAG), Polyakov loop (PG) and Laplacian gauge (LG)). 
Physically relevant types of monopoles (LG and MAG) are distinguished by their 
behavior near the deconfining phase transition. For the LG, Abelian projection 
reproduces well the gauge independent monopole structure encoded in an  
auxiliary Higgs field. Density and localization properties of monopoles, 
their non-Abelian action and topological charge are studied. Results are
presented confirming the Abelian dominance with respect to the non-perturbative 
static potential for all gauges considered.
\end{abstract}

\newpage

\section{Introduction}
Over the last two decades various attempts
have been aiming at a qualitative understanding and modeling
of two basic properties of QCD: quark confinement and
chiral symmetry breaking. Two well-known, complementary
schemes are the instanton liquid model \cite{ILM} and the dual
superconductor picture of the QCD vacuum \cite{DSC1,DSC2}.
While the first model explains chiral symmetry breaking and solves
the $U_A(1)$ problem, the second provides the language to discuss
and to quantitatively describe the confinement mechanism
within an effective field theory. 
Nowadays, it has become of central importance in the lattice community to
quantify the properties of the lattice vacuum guided by the
instanton liquid picture \cite{NEGELE98}. The evidence for a particular
instanton structure is still controversial, in pure Yang-Mills theory 
and in full QCD.

In the effective dual Abelian Higgs theory \cite{ADHM}, the vacuum is viewed
as a dual superconductor, where condensation of color magnetic monopoles
\cite{SmitVdS} (described by the Higgs phase of that model) leads to confinement
of color charges by flux tubes through a dual Meissner effect.
In order to establish this correspondence step by step, it has first
been demonstrated that the Abelian
components of the gauge fields are effective degrees
of freedom (among the original non-Abelian gauge fields) at large distances,
leading to the concept of Abelian dominance \cite{SuzYot}.
The role of monopoles and their different percolation properties for confinement
and deconfinement were empirically substantiated by a large number of lattice
simulations over the last years. It was shown that in the confinement phase
monopoles percolate through the $4D$ volume \cite{BORN} and are responsible for 
the dominant contribution to the string tension (monopole dominance) 
\cite{ShiSuz,BALI}.  More recently, these monopoles have
been shown to be predominantly selfdual (in confinement) \cite{SchiBo} and to
encode the non-perturbative structure of the gauge field at large distances
as seen by quarks, explaining the chiral condensate \cite{MIYAMURA} as
well as the topological density \cite{SasaMiya}.
In order to establish the link to the effective dual theory, the action
that describes the first quantized theory of monopoles has been derived from
the configurations of magnetic currents \cite{SuzAct}. 
At the same time, inherited from the original gauge fields, these ``Abelian
monopoles" are characterized by a self-action (which is essentially
non-Abelian!) and a complicated interaction with the remaining
degrees of freedom.
In the last couple of years also a deeper connection between
monopoles and instantons has been pointed out, both on the lattice
\cite{THUR95,THUR96,SUGA97} and in the continuum
\cite{CHER95,REIN97,WIPF,LENZ,BROW97}.

Following 't Hooft \cite{DSC1}, monopoles should be searched for
as pointlike singularities of some gauge transformation dictated by a local,
gauge covariant composite field. The choice of this field implies the
choice of a particular gauge to analyze the non-Abelian gauge field.
The best known examples are the maximally Abelian gauge (MAG) \cite{KRON87}
and the Polyakov gauge (PG). It has become practice to identify the
monopoles as particle-like singularities on the lattice after
Abelian projection from one of the gauges chosen, identifying their
trajectories in the resulting compact Abelian
lattice field. This is known as the DeGrand-Toussaint (DGT) construction
\cite{degrandT}.
An attractive alternative, related to
the Laplacian gauge (LG), has been revived recently \cite{VINK,SIJS}.
This method offers the possibility to identify monopoles on the lattice
without the problems that the conventional methods have and bears
a close relationship to the 't Hooft-Polyakov monopoles.
One can define them without really performing a gauge transformation.
The LG can, however, also be the starting point for an
Abelian projection followed by a DGT monopole construction.
In this paper we intend to compare these three gauges used to define
DGT monopoles with respect to their  physical significance  
at the deconfinement transition.
Furthermore we want to give a justification of the DGT construction 
of monopoles in the case of the LG, in view of their gauge 
independent localization similar to that of 't Hooft-Polyakov monopoles.

An important ingredient of our discussion
is a smoothing method for non-Abelian gauge fields used to remove UV
fluctuations that are without relevance for the long distance physics.
From our previous experience \cite{FEUE} with this method we know that
this procedure is suitable to eliminate UV lattice artifacts from
the monopole-related observables, too, without destroying the confining
structure as usual cooling procedures finally do.
Originally, methods like cooling or smoothing have been developed in
order to study the instanton structure which otherwise is possible only by
fermionic (spectral flow) methods \cite{fermionic}.
A particular feature of cooling or smoothing, that should interest us here,
is that it maps gauge fields on full gauge field configurations which 
still can then be put into various gauges for a closer study 
of one or the other qualitative picture.  
Concerning the monopole degrees of freedom, it is advantageous that smoothed 
configurations carry monopole currents with short loops removed.

The renormalization group (RG) based smoothing method which uses
(classically) perfect actions has been strongly recommended in the
last two years \cite{degrand,FEUE,PRD98} because it is not afflicted
with some problems of the simple or improved cooling \cite{coolimp} methods.
UV fluctuations near to the lattice discretization scale are removed
from the gauge field by one blocking step followed by a smooth interpolation
of the field on the original lattice. By this procedure, 
performed locally below a well-defined (the blocked) scale, the RG smoothed 
configurations remain confining (if they are derived from MC runs in the 
confining phase). The string tension is reproduced almost completely. 
At the lattice sizes (and values of $\beta$) that were to our disposal, 
the two RG levels involved in the procedure are not totally decoupled from 
the confinement scale. Fluctuations are removed by the 
blocking--inverse--blocking scheme, which contribute a few percent to the string 
tensions at the distances where we can study it.

In this paper we put special emphasis on the possibility of a gauge invariant
definition of monopole trajectories and test it against the conventional
DGT method usually applied to configurations being Abelian projected from various 
gauges. We were particularly interested in temperatures below and above the 
deconfinement critical point. This gave  us the opportunity to compare the 
behavior of the respective monopole degrees of freedom 
at the transition and examine their dynamical relevance.  

The paper is organized as follows.
In Sec. II we briefly discuss the RG smoothing technique and recall some
properties of action and topological charge densities after smoothing.
In Sec. III we describe the different gauge fixing procedures being used and
report certain technical details how they work for smoothed configurations.  
In Sec. IV  various aspects of the monopole degrees of freedom are presented, 
pinning down the main difference between the different gauges  
by means of the different physical behavior of the corresponding 
monopoles across the phase transition.  In Sec. V we 
present some incidental results of our study concerning Abelian dominance, 
comparing the Abelian string tensions obtained by Abelian projection from 
various gauges with the fully non-Abelian non-perturbative potential extracted
by smoothing.  In Sec. VI we summarize our conclusions.

\section{RG Smoothing and Physical Densities
\label{localact}}

To recognize semiclassical structure in ensembles of
gauge field configurations generated  by lattice simulations,
cooling methods have been used, that minimize the action.
However, even improved versions of cooling \cite{coolimp}
gradually reduce the string tension. These versions are designed
to stabilize instantons above a certain threshold
for their scale size $~\rho~$ (typically $~\rho > 2a~$
with the lattice spacing $a$).  However, they let close 
instanton-antiinstanton pairs annihilate.

Up to now, most lattice studies are performed using the Wilson action,
even if they focus on the topological structure of the vacuum.
This action is known to be afflicted with so-called dislocations,
structures of a size near one lattice spacing.
For dislocations certain lattice definitions of the topological charge $Q$ give
a signal of unit charge, although their total action $S$ violates the inequality
$S \geq |Q| 8\pi^2 / g^2$ and the entropic bound $S > \bar{S}=48\pi^2/11~N_c^2$.
For discretized instantons, the Wilson action decreases with
size $\rho$ of an instanton, such that isolated instantons shrink under cooling 
before they decay as a dislocation.
Thus, cooling with Wilson action is unsuitable to unambiguously determine 
the instanton size. Other methods like APE smearing let instantons even grow.

To avoid these ambiguities we have proposed to use the renormalization group
motivated method of ``constrained smoothing'' \cite{FEUE} which is based
on the concept of perfect actions \cite{HASE}.
For discretized instantons, these actions guarantee a size independent
action for instantons, even with radii near to the lattice spacing.
They allow a theoretically consistent ``inverse blocking''
operation which, for instance, reconstructs an instanton solution
on a finer lattice with the same value of action.
Inverse blocking is a method to find a smooth interpolating field on a
fine lattice by constrained minimization of the perfect action, provided
the configuration is given on a coarse lattice. This way to interpolate
allows to define an unambiguous topological charge \cite{HASE}.

For non-classical configurations like generic Monte Carlo configurations,
one is interested to study smoothed configurations which are
made free from UV fluctuations below a well-defined scale.
Constrained smoothing in the way we use it, consists in
first blocking the fields $\{U\}$, sampled on a fine lattice with lattice
spacing $a$ using the perfect action, 
to a coarse lattice configuration $\{V\}$ with lattice spacing $2~a$
by a standard blockspin transformation.  Then inverse blocking is used to
find a smoothed field $\{U^{\mathrm{sm}}\}$ replacing $\{U\}$ on the
fine lattice.  We refer to this method as ``renormalization group'' (RG)
smoothing.

This procedure can be cyclically repeated, choosing different blocked 
lattices for the blocking step in subsequent cycles. This has been
practized in Ref. \cite{boulder_cycling}. Experience shows that this RG 
cycling method makes configurations locally more and more 
classical. Unfortunately, 
there is no well-defined smoothing scale, but in contrast to cooling 
this method preserves long range features like confinement rather well.
An important advantage of the simpler RG smoothing method outlined
above is that it
does not drive configurations into classical fields as unconstrained 
minimization of the action (and RG cycling) would do. It saves the 
long-range structure of the Monte Carlo configuration in $\{V\}$
and allows to study semiclassical objects in the smoothed background
deformed by classical and quantum interaction. It is the upper blocking 
scale which roughly defines the border line between ``long'' and
``short range''.

In this work we used a simplified fixed-point action \cite{degrand,PRD98} 
for Monte Carlo sampling and for RG smoothing.
Before describing in detail the various gauge conditions we should
explain which {\it gauge invariant} quantities we shall use to characterize
the smoothed configurations. Partly they are suggested by the perfect
action $S_{FP}$ itself which is parametrized in terms of only two types
of Wilson loops, plaquettes $U_{C_{1}}=U_{x,\mu,\nu}$ (type $C_{1}$)
and tilted $3$-dimensional $6$-link loops (type $C_{2}$) of the form
\begin{equation}\label{sixlinks}
U_{C_{2}}=U_{x,\mu,\nu,\lambda}=
U_{x,\mu}
U_{x+\hat{\mu},\nu}
U_{x+\hat{\mu}+\hat{\nu},\lambda}
U_{x+\hat{\nu}+\hat{\lambda},\mu}^{\dag}
U_{x+\hat{\lambda},\nu}^{\dag}
U_{x,\lambda}^{\dag} \, \, ,
\end{equation}
and contains several powers of the linear action terms
corresponding to each loop of both types that can be drawn on the lattice
\begin{equation}\label{eq:fp_action}
S_{FP}(U)=\sum_{type~ i} \sum_{C_{i}} \sum_{j=1}^{4}
w(i,j) (1 - {1 \over 2}~\mbox{Tr}~U_{C_{i}} )^j  \, \, .
\end{equation}
The parameters of this action are reproduced in Table \ref{tab:weights}.

We define the action density $s_{site}(x)$ per lattice point
(a local action) as follows
\begin{equation}
\label{eq:local_action}
s_{site}(x) =  \sum_{j=1}^{4}
\left(
  \sum_{C_{1}(x)}  \frac{w(1,j)}{4} (1 - {1 \over 2}~\mbox{Tr}~U_{C_{1}(x)})^j
+ \sum_{C_{2}(x)}  \frac{w(2,j)}{6} (1 - {1 \over 2}~\mbox{Tr}~U_{C_{2}(x)})^j
\right)
 \,.
\end{equation}
Here, $~C_{j}(x)$ ($j=1,2$) means loops of type $j$ running through
the lattice site $x$.  Summing $s_{site}(x)$ with respect to $x$ yields the 
total action of the configuration.

The topological charge density definition we are using is ({\it i})
the simplest ``field theoretic'' one constructed out of plaquettes around 
a site $x$
\begin{equation}\label{eq:q_naive}
q(x) = - \frac1{2^9 \pi^2} \sum_{\mu,\nu,\sigma,\rho=-4}^{+4}
\epsilon_{\mu\nu\sigma\rho}
{\mathrm tr} \left(U_{x,\mu,\nu} U_{x,\sigma,\rho}\right)
\end{equation}
and ({\it ii}) the charge contributed by hypercubes
according to L\"uscher's definition of topological charge \cite{luescher}.
We did not attempt to improve the field theoretic definition.
Both topological densities used are known to behave regularly
for smooth configurations \cite{letter96,FEUE}.

In order to  give a characterization of Abelian monopole currents
in terms of locally defined {\it gauge independent} observables,
we consider also the  $3$-cubes of the original lattice dual
to an elementary piece of monopole world line (carried by a link of the 
dual lattice).  This leads us to a natural definition
of a local action on the monopole world line, $s_{3-cube}(c)$
instead of $s_{site}(x)$.
We include into $s_{3-cube}(c)$ the contribution of all plaquettes forming
the $6$ faces of the $3$-cube $c$ plus the contribution of all $6$-link
loops which wind around its surface. In short, it is the part of the
total action which lives on that cube. 

\section{Monopole Identification Methods}
\subsection{Maximum Abelian gauge and Laplacian gauge}
The most popular gauge is the maximally Abelian gauge (MAG) \cite{KRON87}.
Abelian projection of configurations put into MAG exhibits Abelian dominance.
This gauge is often used to study the dynamics of monopoles on the lattice. 
It is enforced by an iterative minimization procedure, which can get stuck 
in local minima that are gauge copies of each other (so-called {\it technical} 
Gribov copies). The Laplacian gauge (LG) is not afflicted with this problem.
The basic idea has been suggested \cite{Seixas} more than a decade 
ago but has apparently not been used until very recently \cite{SIJS}.
Originally, in the context of the adjoint Higgs (Georgi-Glashow) model, 
it has been proposed to introduce, besides the quantized scalar matter field, 
a purely auxiliary adjoint slave field.

Such a Higgs field can be used to define a gauge transformation
to the ``unitary'' gauge with respect to the auxiliary field except at
its zeroes.
Thus, the corresponding monopoles are located at the zeroes of the 
auxiliary Higgs field and are representing the singularities of the 
gauge rotation. They can be identified as t'Hooft-Polyakov monopoles.
For each given (generic, non-classical) non-Abelian gauge field the
auxiliary field is determined as the eigenvector related to the
lowest eigenvalue of the adjoint covariant Laplacian. This is why the 
resulting gauge is named Laplacian gauge (LG).  
The monopole identification via the Higgs zeroes is obviously gauge invariant 
(see also Ref. \cite{HOLLANDS}).
It is this prescription we want to apply to pure Yang-Mills theory
on the lattice and to compare with other ways for the Abelian
projection. 
Strictly speaking, in order to describe the location of the monopole
world lines (Higgs zeros), we would need algorithms interpolating the Higgs
field between the lattice sites. For our present purposes it will be
sufficient to detect clusters of lattice points with small Higgs modulus.

Technically, MAG and LG are closely related to each other.
For MAG, the gauge functional to be minimized can be written as follows
\begin{eqnarray}
\label{functional}
F(\Omega) & = &\sum_{x,\mu} (1-\frac{1}{2} \, tr \,
          (\sigma_3 U^{(\Omega)}_{x,\mu}
           \sigma_3 U^{(\Omega)\dag}_{x,\mu} ) ) \nonumber \\
&=&  \sum_{x,\mu,a} (X_x^a - \sum_{b} R^{a,b}_{x,\mu}
          X_{x+\hat \mu}^b )^2 \rightarrow \int_V (D_{\mu} X)^2 \, ,
\end{eqnarray}
with the gauge transformation $\Omega_x$ acting on $\{U\}$
\begin{displaymath}
U^{(\Omega)}_{x,\mu}  = \Omega_x U_{x,\mu} \Omega^{\dag}_{x+\hat \mu}
\end{displaymath}
encoded in an {\it auxiliary} adjoint Higgs field
\begin{displaymath}
\Phi_x  =  \Omega^{\dag}_x \sigma_3 \Omega_x = \sum_a X^a_x \sigma_a
\end{displaymath}
subject to local constraints $\sum_a (X^a_x)^2 =  1$
and with adjoint links
\begin{displaymath}
R^{a,b}_{x,\mu}  =  \frac{1}{2} \, tr \, (\sigma_a U_{x,\mu} \sigma_b
                                               U_{x,\mu}^{\dag}) \, .
\end{displaymath}
To change from MAG to LG, the local constraints
$ \sum_{a} (X^a_x)^2 = 1 $
are relaxed and replaced by a global normalization:
$ \sum_{x,a} (X^a_x)^2 = V $, such that
Eq. (\ref{functional})
can be further written:
\begin{equation}
\int_V (D_{\mu} X)^2 \rightarrow \sum_{x,a} \sum_{y,b}
X^a_x\{-\Box_{x,y}^{a,b}(R) \} X^b_y \, .
\end{equation}
Thus, the minimization of the gauge functional is reduced to a search for the 
lowest eigenmode of the covariant lattice Laplacian. The LG of a given
lattice configuration is unambiguously defined, except for the case that
degenerate lowest eigenmodes exist. The corresponding LG configurations
would then define {\it true} Gribov copies of each other.
For both MAG and LG, the gauge transformation is finally accomplished 
by finding $\Omega_x$ that diagonalizes the field $\Phi_x$.

\begin{figure}[h]
\epsfxsize=15.0cm\epsffile{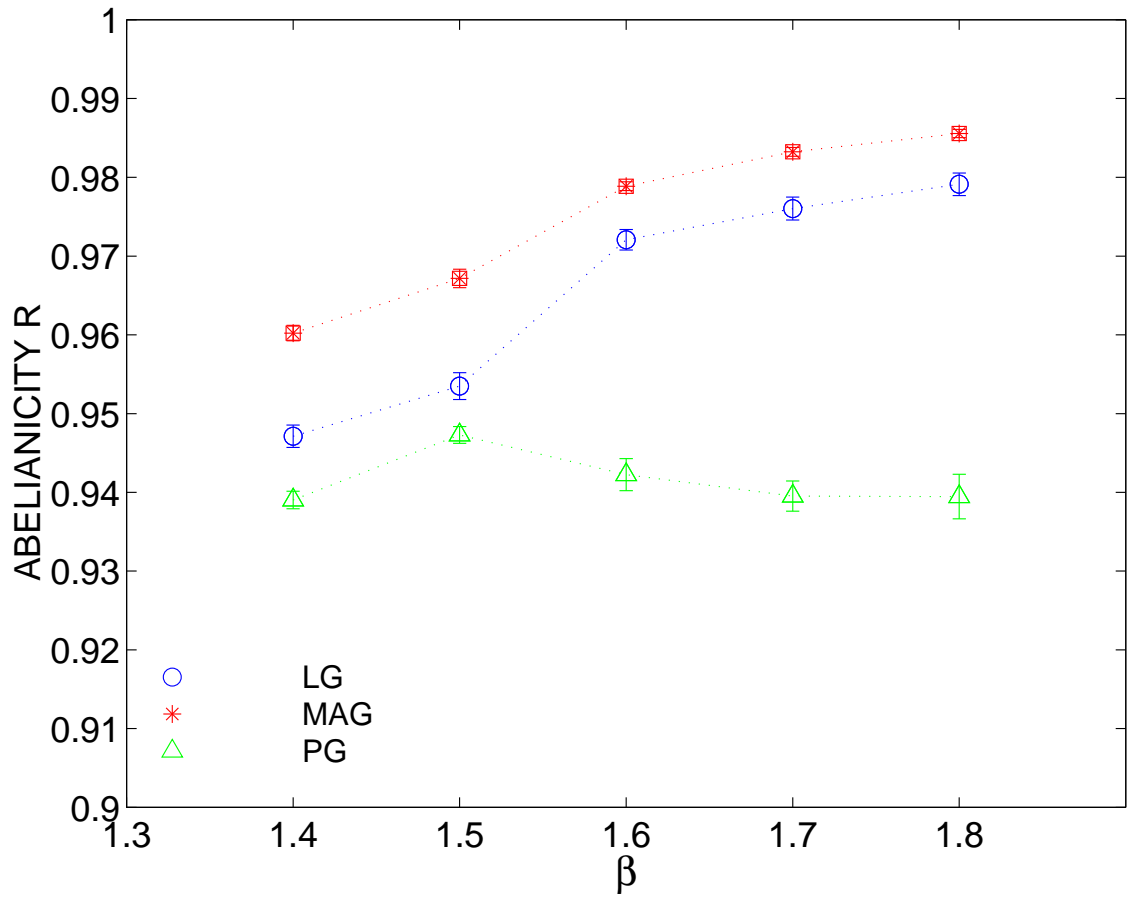}
\caption{
      Abelianicity as a function of $\beta$ for MAG, LG and PG.
\label{fig:abelianicity}
}
\end{figure}

Here a technical observation ought to be made. As usual, the MAG is found 
iteratively by minimizing the functional (\ref{functional}).
The fact that we analyze smoothed configurations does not mean that
the MAG is computationally less demanding to find. Still, as in the case for 
unsmoothed configurations, we need $O(1000)$ iterations to fulfill our
stopping criterion. We require  that the next gauge 
transformation should deviate from unity uniformly by less than $10^{-7}$.
By abelianicity $R$ we denote the average Abelian fraction per link
after completion of the minimization.  We show the $\beta$ dependence of this
quantity in Fig. \ref{fig:abelianicity} in comparison with other
gauges which do not explicitly attempt to get a large abelianicity.
There is a monotonous increase with $\beta$ across the phase transition, 
but no ``deconfinement signal'' related to it.

\begin{figure}[h]
\epsfxsize=15.0cm\epsffile{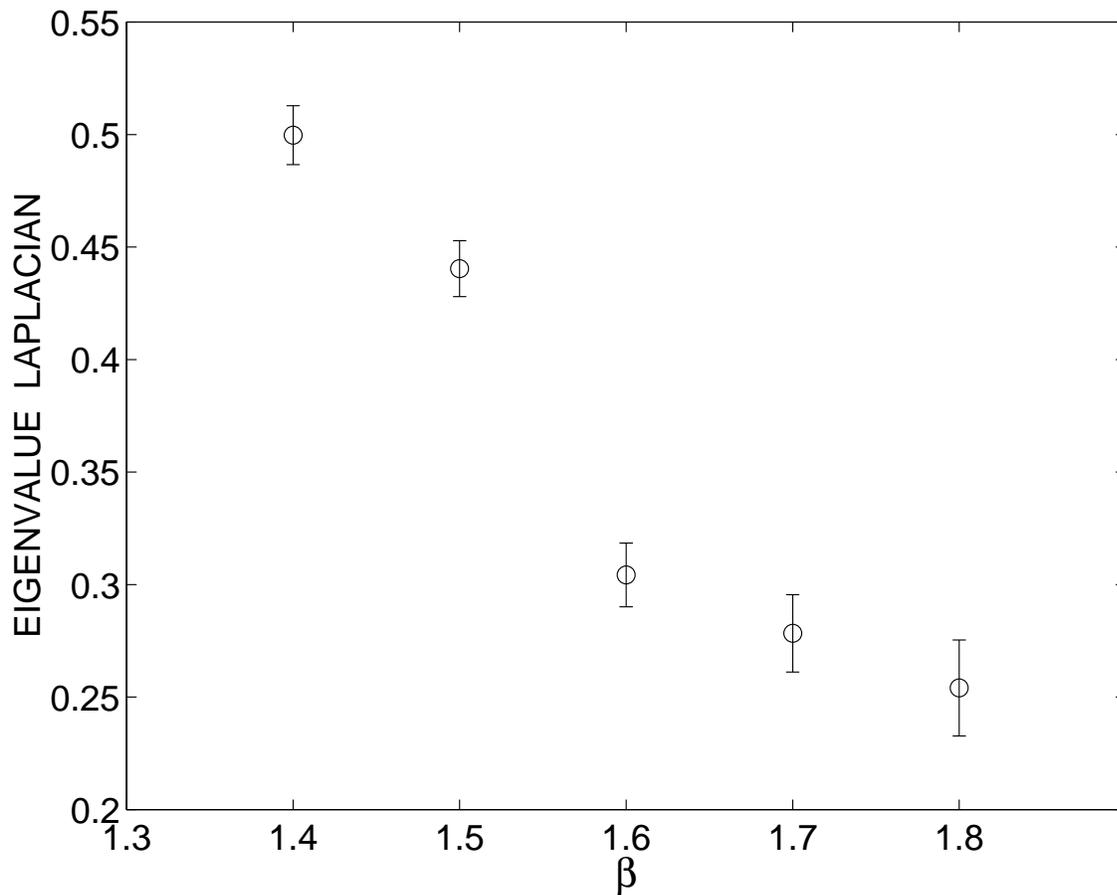}
\caption{
      Average over 100 independent configurations 
		of the lowest eigenvalue of the adjoint
      lattice Laplacian on the $12^3 \times 4$ lattice 
      as a function of $\beta$ or temperature.
\label{fig:gaugefixing_lap}
}
\end{figure}

For the search of the lowest eigenvalue for the LG and its corresponding 
eigenvector, we apply the conjugate gradient algorithm to minimize the following
Ritz functional
\begin{equation}
F=\frac{\langle X ,( - D^2 ) X \rangle}{\langle X , X \rangle} \, ,
\end{equation}
where $D^2$ is the adjoint lattice Laplacian. 
We choose 5 random start vectors $X_x=X^a_x \sigma_a$  and follow the 
relaxation of the estimated eigenvalue over 3000 iterations.
The vector with the lowest eigenvalue is then taken further representing 
the given gauge field configuration. The average over an ensemble of 100 
gauge fields (per $\beta$ value) of the respective lowest eigenvalue
is shown in Fig. \ref{fig:gaugefixing_lap} as a function of $\beta$ 
(or temperature). This average is $\beta$ dependent in both phases with a 
steep drop across the transition.
The corresponding abelianicity in Fig. \ref{fig:abelianicity} 
is always smaller than the abelianicity for
MAG, but approaches it with increasing $\beta$ (in the deconfined phase).  
Among the different gauges, the biggest rise of abelianicity accompanying 
the transition is found in the LG case.

\subsection{Polyakov gauge}
This gauge condition puts special emphasis on the 
(gauge group valued) Polyakov loop $P(x)$, 
the gauge transporter along the shortest loop starting from and arriving at a 
given lattice point which is closed by periodic thermal boundary conditions.
The Polyakov gauge is enforced by simultaneous diagonalization of $P(x)$
at each point.  In Fig. \ref{fig:abelianicity} we also show the abelianicity
achieved in the PG. Except near the transition (where it has a shallow maximum)
the abelianicity is practically $\beta$ (temperature) independent.
This gauge has the smallest abelianicity among the discussed
gauges, but it might be surprising that it comes near to $95$ \%.

\subsection{Monopole identification}
After the gauge of choice has been fixed one extracts the Abelian
degrees of freedom by Abelian projection. The Abelian link angles
representing a $U(1)$ field with residual gauge symmetry
can then be used for the identification of monopoles. This is done,
like in pure compact $U(1)$ gauge field theory,
by computing the $U(1)$ gauge invariant (magnetic) fluxes through the
surfaces of elementary 3-dimensional cubes or -- what is equivalent --
by searching for the ends of Dirac strings.
Monopoles identified in this manner are generally referred
to as DeGrand-Toussaint (DGT) monopoles \cite{degrandT}.
We are going to localize DGT monopoles in our smoothed configurations
put into all three gauges and to compare their possible significance.

As stressed before, for the LG there exists an independent, gauge invariant 
way to localize monopoles in the sense of 't Hooft 
that can be used to test the reliability of the DGT method.
The Higgs field introduced in the LG provides an alternative,
physically more satisfactory possibility of monopole identification.
In the continuum, lines of $\rho_x=|\Phi_x|=0$,  where $\rho_x$ is defined as
\begin{equation}
 X^a_x = \rho_x \hat X^a_x \,\,\ , \,\,\,
\rho_x = \sqrt{\sum_{a=1}^3 (X^a_x)^2 }  \, ,
\end{equation}
directly define lines of gauge fixing singularities (mo\-no\-po\-les).
For the 't Hooft-Polyakov monopole in the adjoint Higgs model, regions with 
$\rho=0$ of the {\it physical} Higgs field are identified
with the centers of such monopoles. Note that this way of monopole
identification in the pure Yang-Mills theory, too, does not require to 
actually perform the gauge transformation and an Abelian projection!
It is gauge invariant by definition.

\begin{figure}[h]
\begin{tabular}{c c }
\epsfxsize=7.4cm\epsffile{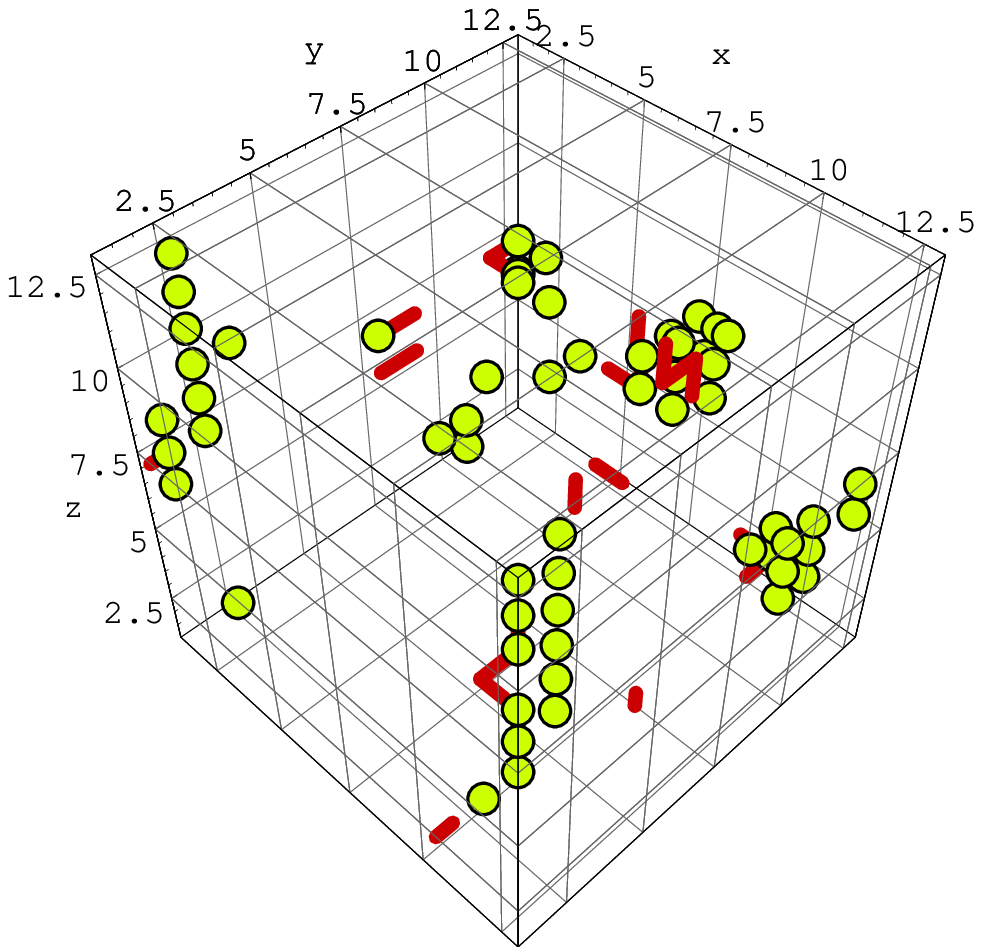} &
\epsfxsize=7.4cm\epsffile{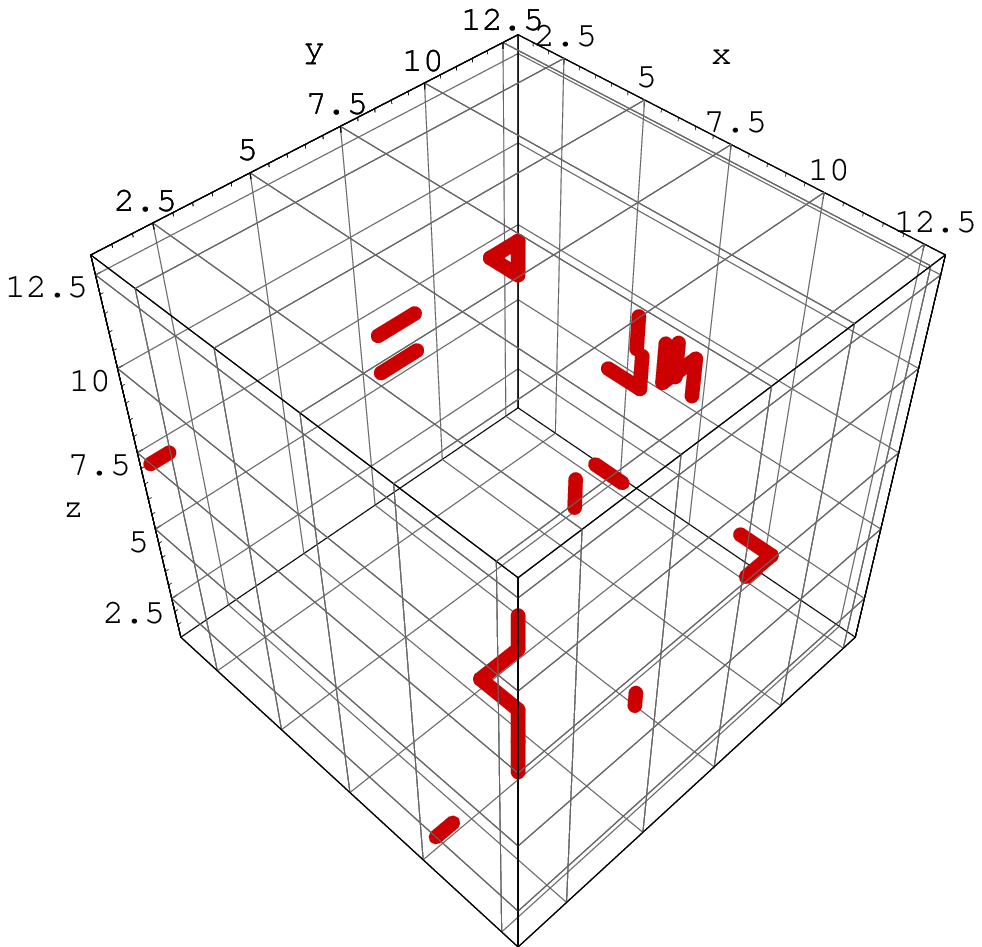} \\
\epsfxsize=7.4cm\epsffile{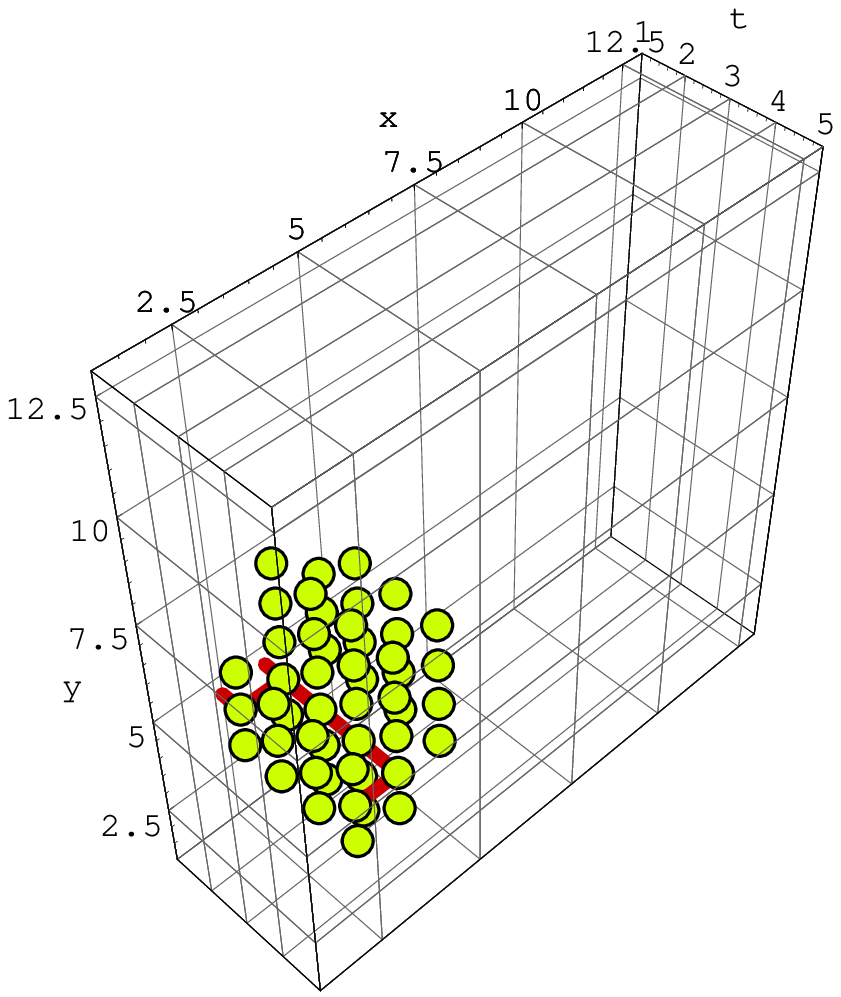} &
\epsfxsize=7.4cm\epsffile{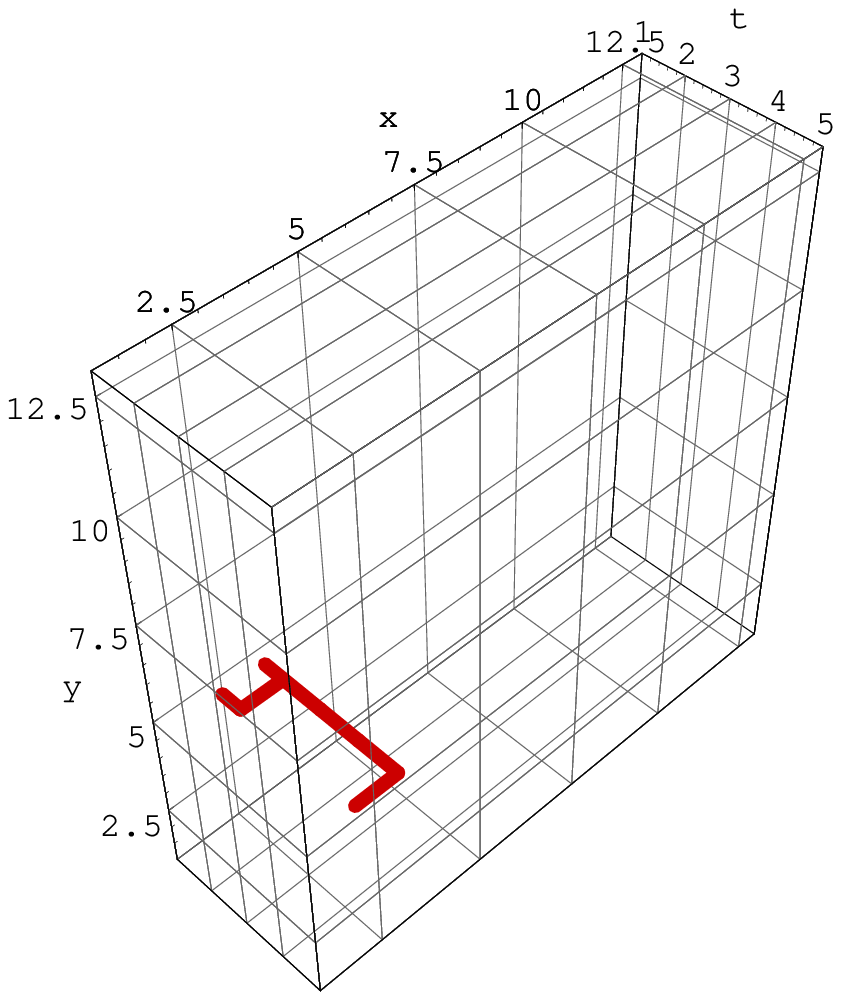} \\
\end{tabular}
\caption{ 
      Regions with local Higgs norm $\sum_a (X_x^a)^2 < 10^{-5}$ 
      (marked by dots) of the lowest eigenmode $X_x^a$ of the adjoint 
      lattice Laplacian (normalized over the $12^3 \times 4$ lattice).
      Monopole loops (shown as lines, for clarity shown also separately in 
      the right plots) have been identified following DGT after Abelian 
      projection from the Laplacian gauge. The upper configuration is taken 
      from the confinement ensemble for a fixed time slice, 
		the lower from the deconfined phase for $z$ fixed.
\label{fig:Xlokal}
}
\end{figure}

\begin{figure}[h]
\epsfxsize=14.5cm\epsffile{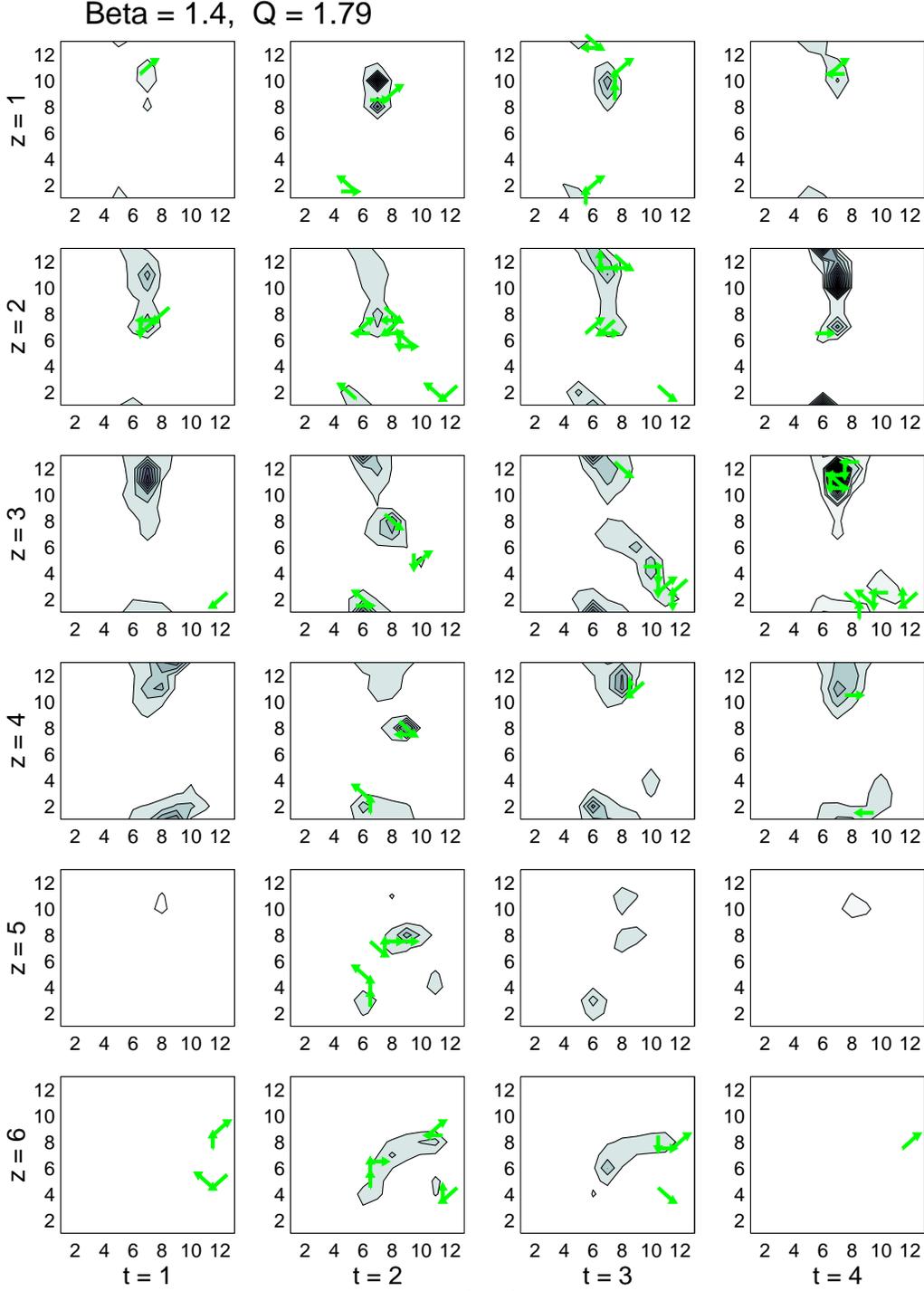}
\caption{ 
      Contour plots in the $(x,y)$-plane of the local Higgs norm (shown only for  
      $\sum_a (X_x^a)^2< 10^{-5}$) in the $z=1,..,6$ half of the lattice 
      for all time slices $t=1,..,4$. Magnetic currents of DGT monopoles 
      as obtained by Abelian projection from LG are symbolized by  arrows
		with non-vertical or non-horizontal directions indicating 
      the $z$ and $t$ directions.
\label{fig:schicht}
}
\end{figure}

Let us demonstrate the localization of the singularities 
of the LG transformation 
and how this is related to the DGT monopoles. We have plotted
in Fig. \ref{fig:Xlokal} all lattice points with the local modulus
squared $X^2(x)=\sum_a (X_x^a)^2$ being less than a threshold value of $10^{-5}$,
for a generic configuration
of the confined phase ($t$ fixed, top) and for one of the 
deconfined phase ($z$ fixed, bottom).
The eigenvectors are always globally (over $12^3 \times 4$ lattice points)
normalized to one, $\sum_x \sum_a (X_x^a)^2 = 1$.
Although there are no true singularities (exact zero modulus of $X$) located
on lattice sites
it can be seen that small values $X^2(x)$ (less than one order of magnitude
below the average) mark connected regions of space
on the lattice where one may expect to find the monopole world lines
(see also Ref. \cite{SIJS}). The dark lines correspond to DGT monopoles
that have been extracted after Abelian projection has been applied to the LG.
These monopoles are shown for a confinement configuration 
in Fig. \ref{fig:schicht} by arrows, where  
non-vertical and non-horizontal directions encode world lines leaving (entering)
each $(x,y)$-miniplot along the $z$ or $t$ direction.
They are located almost completely inside the shaded regions having 
$X^2(x)<10^{-5}$.

Due to the smoothness property of the gauge transformations dictated by the
local field $X_x^a \sigma_a$ it has been conjectured in Ref. \cite{SIJS}
and demonstrated for normal Monte Carlo configurations that monopole
currents are more abundant in the LG compared with MAG, as far as DGT
monopoles are concerned. For RG smoothed gauge field configurations, we
confirm this conjecture to be true as well.

\begin{figure}[h]
\begin{tabular}{c}
\epsfxsize=12.0cm\epsffile{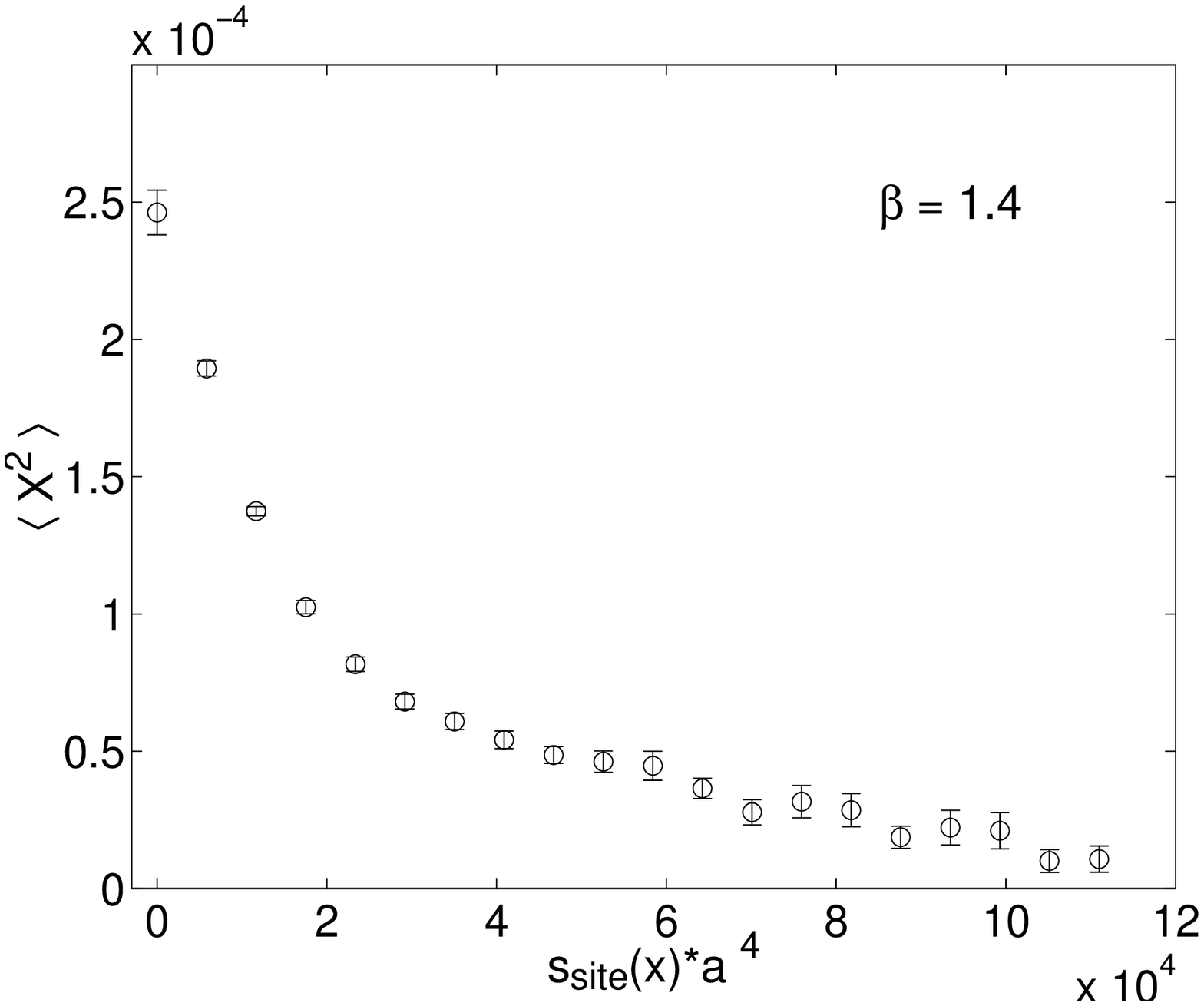} \\
\epsfxsize=12.0cm\epsffile{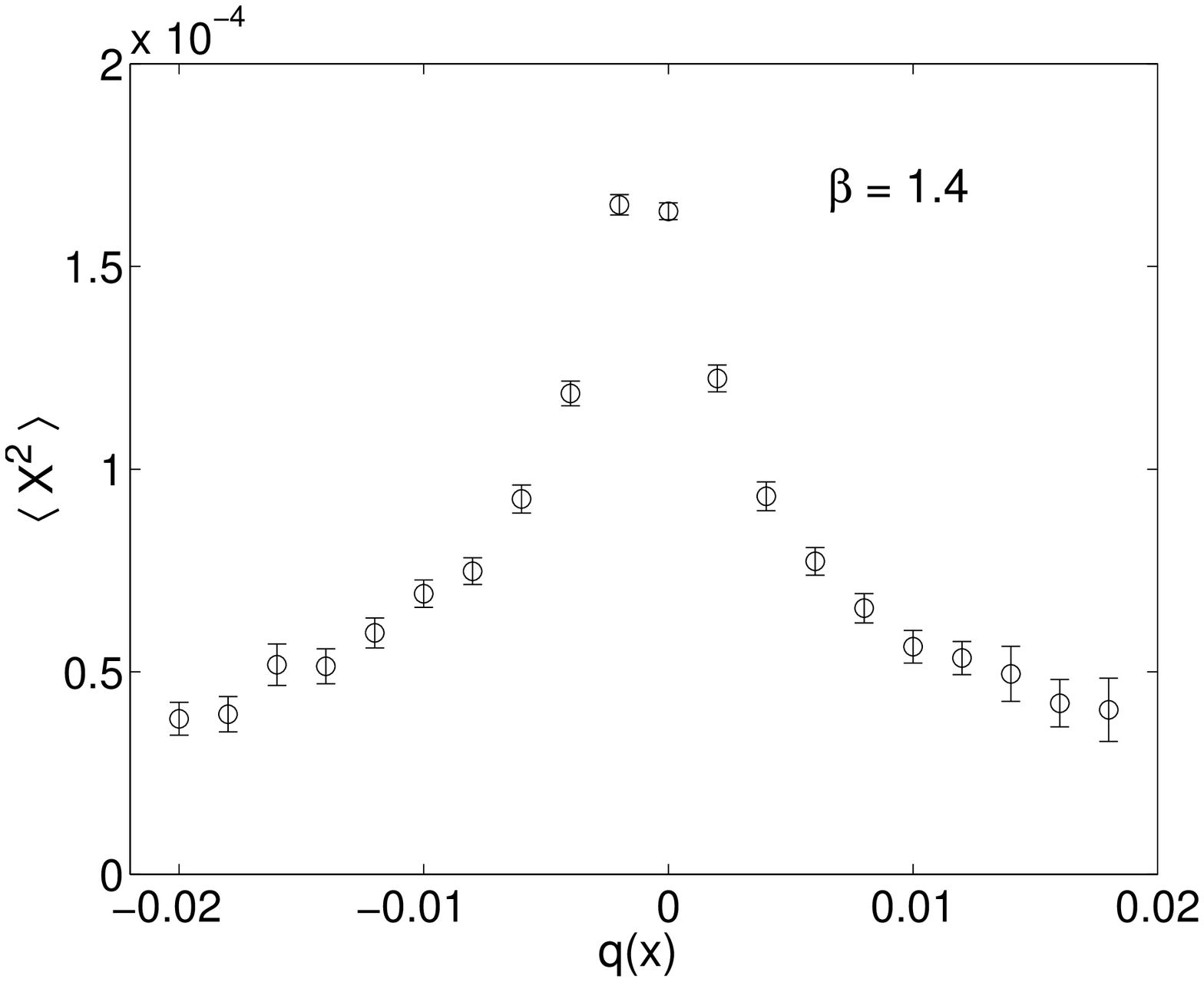}
\end{tabular}
\caption{
      Average of the local Higgs 
      norm $X^2(x)=\sum_a (X_x^a)^2$ for lattice points
      having a value of local action $s_{site}$ within the same respective 
      bin (top). The same for lattice points binned according to topological 
      charge density $q$ (bottom).
\label{fig:binX}
}
\end{figure}

A small local modulus squared $X^2(x)=\sum_a (X_x^a)^2$
expresses the obstruction to unwind
the gauge field (to the unitary gauge) near the site $x$, while the local
contribution to the Ritz functional $\sum_a X^a_x ( - D^2 X)^a_x$ 
expresses the disordering of the gauge field.
Here we concentrate on the Higgs modulus squared because this 
is directly related to the monopole localization.
In Fig. \ref{fig:binX} the upper plot shows the average value of $X^2$ provided,
site by site, the
local action $s_{site}$ falls into one of the bins on the abscissa.
This plot refers to $\beta=1.4$.  It lets us expect that
only points with $s_{site}>4 a^{-4}$ have a
chance to meet the condition $X^2<10^{-5}$ mentioned before with
non-vanishing probabiltity.
In Fig. \ref{fig:binX} the lower plot (also for $\beta=1.4$) shows
the average value of $X^2$ similarly depending on
the local topological density $q$ falling into one of the bins.
According to this, only points with $|q|>0.01$ have a chance to meet the
$X^2$ condition mentioned before with some probability.

\begin{figure}
\begin{tabular}{cc}
$\beta=1.4$ & $\beta=1.8$ \\
\epsfxsize= 7.5cm\epsffile{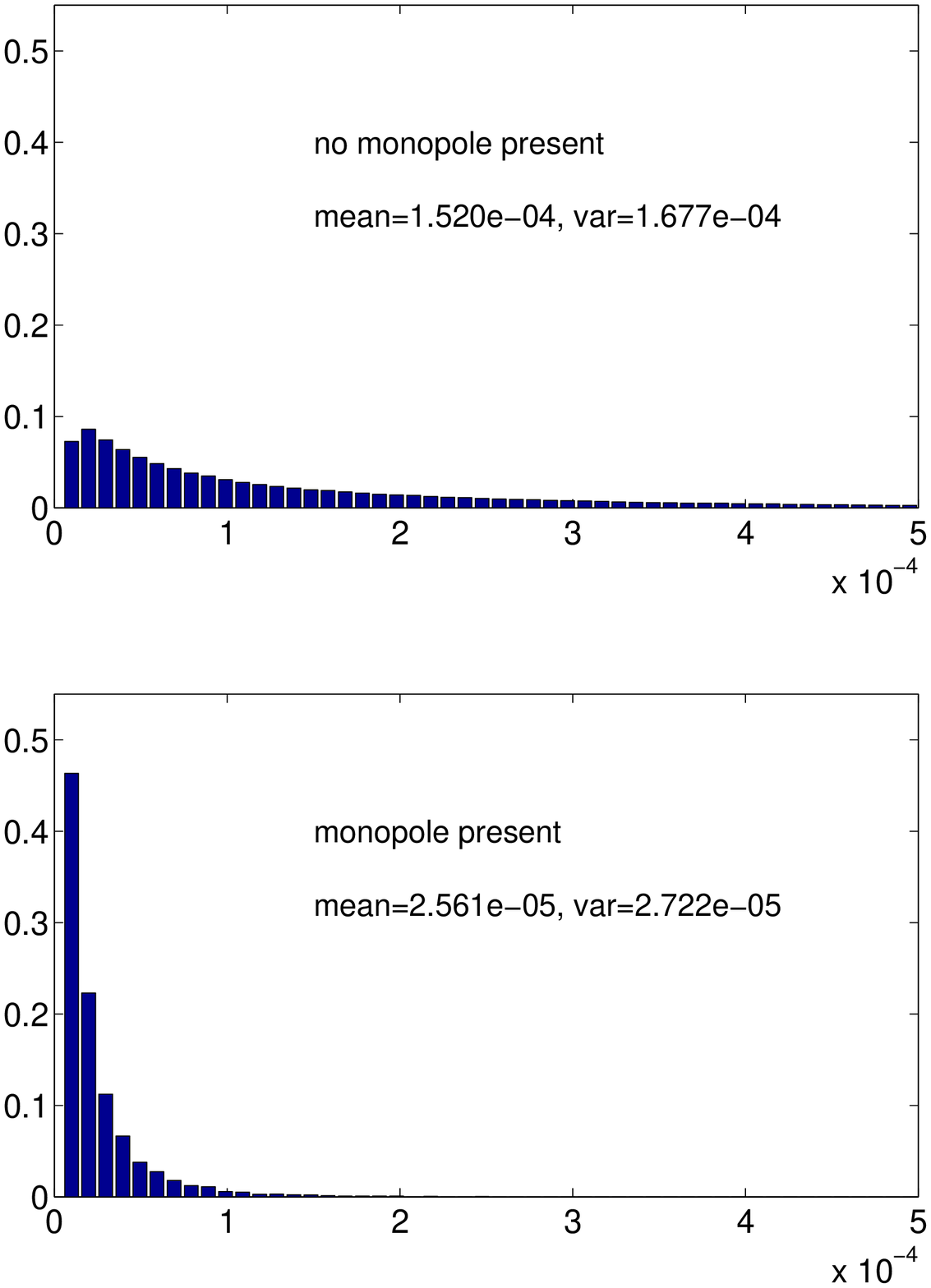} &
\epsfxsize= 7.5cm\epsffile{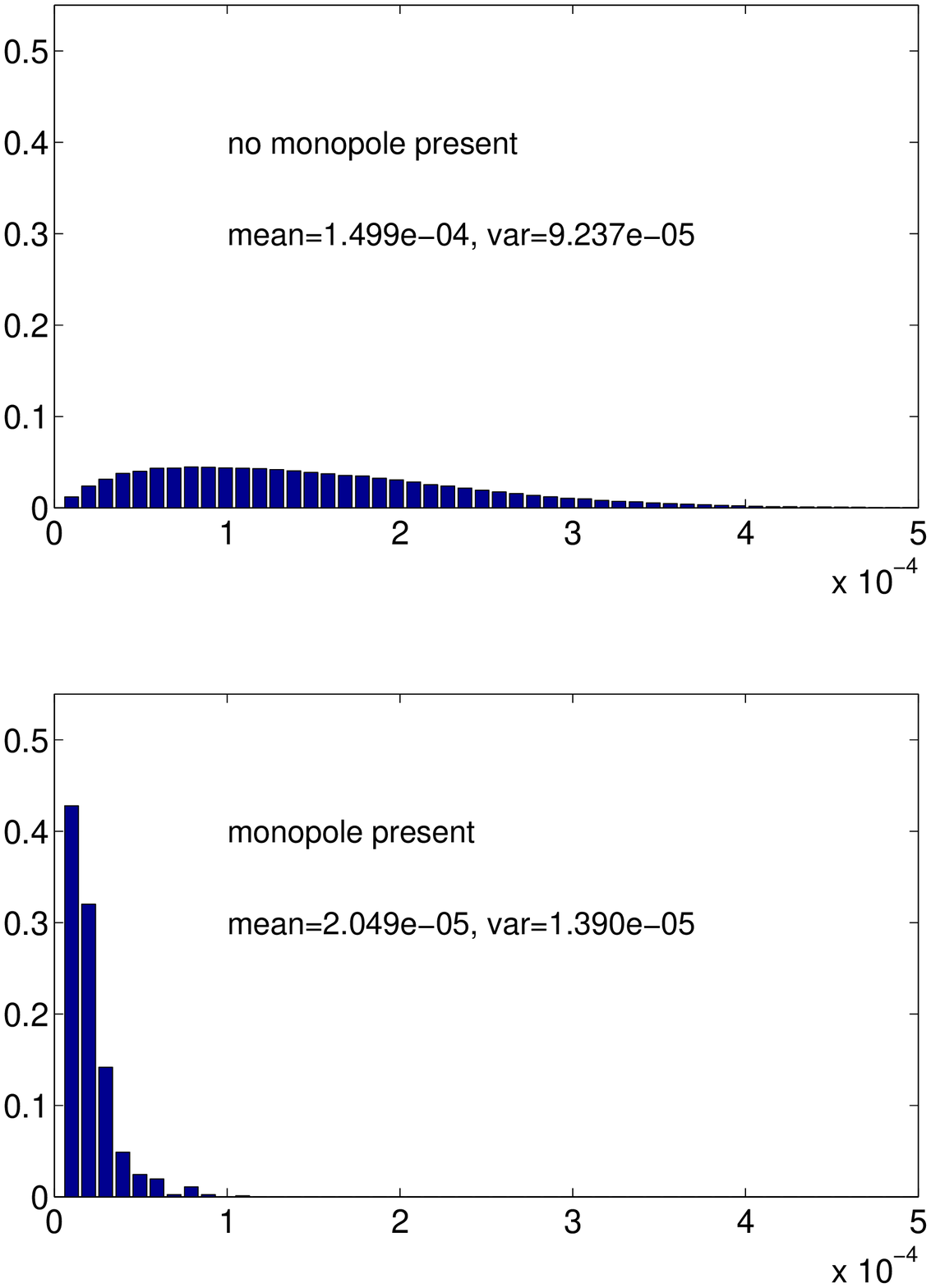}\\
{\small $ X^2$ }  & {\small $ X^2$ }
\end{tabular}
\caption{
      Probability distribution of the local Higgs 
      norm $X^2$ in the absence (top) and 
      in presence (bottom) of a DGT monopole obtained in the LG. 
      $\beta=1.4$ (left) refers to the confinement, $\beta=1.8$ (right) to the 
      deconfinement phase.
\label{fig:probX}
}
\end{figure}

In Fig. \ref{fig:probX} we show the probability distribution
for encountering a certain $X^2$ for the two cases, having no DGT monopole
in a $3D$ box (top) and having one or more monopoles present (bottom), 
both in the confinement and in the deconfinement 
phase. To the (normalized) histograms, values of $X^2$ found at the eight 
corners of each (occupied or empty) $3D$ cube have contributed. The histograms
are averages over all configurations.
It is obvious that in the case of a DGT monopole
the $X^2$ distribution on the corners is strongly concentrated near zero. 
Roughly $50$ \% of the neighboring lattice sites fall below the threshold
mentioned above.

Finally, we should add one remark on the problem of potential Gribov copies.
We have checked that the DGT monopoles, detected on RG smoothed configurations
after transforming to MAG and subsequent Abelian projection, do basically not 
change their positions if random gauge transformations are applied before 
relaxation to MAG.
From several RG smoothed configurations we have produced 10 different randomly
gauged copies and have analyzed the relative shifts of individual monopole
trajectories that emerge after MAG, Abelian projection and DGT construction.
We found that more than $60$ \% of the dual links occupied by monopole
trajectories will stay unchanged in position. About $30$ \% of the links
will be shifted by one lattice spacing, shifts of two and more units were
very rare. This gives reason to believe that RG smoothed 
fields are smooth enough such that the MAG gauge and Abelian projection will 
identify singularities without ambiguity. 
Notice that this was not the case without smoothing \cite{BORN,BALI}.
Concerning the LG method, the eigenvalue estimators slightly differ numerically
according to the 5 different random starting vectors that we considered, 
the regions 
characterized by a Higgs modulus below the threshold do practically not differ 
between the relaxed vectors. We interprete our observations mentioned before 
as an {\it a posteriori} justification for the use of the DGT method
in the case of LG, too. In the following we will exclusively use the DGT
prescription to extract LG monopoles.

\section{Physical results on Abelian monopoles}
Our configurations are generated on a $12^3 \times 4$ lattice at various
$\beta$-values with the perfect action tested for consistency in 
Ref. \cite{PRD98}. In that paper we have found a critical 
$\beta_c=1.545(10)$ of the deconfinement transition for $N_{\tau}=4$. 
According to the second order of the transition, we have determined it
from the intersection of the Binder cumulant of Polyakov lines
(unblocked and blocked ones) for different spatial volumes.
The present values of $\beta=1.4$, $1.5$, $1.6$, $1.7$ and $1.8$
enclose the critical point. The smoothing is performed as described in
Ref. \cite{PRD98}.
We have extracted the monopole degrees of freedom in the respective
gauge by projection onto the diagonal part of the $SU(2)$ gauge field.
For this Abelian gauge field, we have then detected the DGT monopoles.

\subsection{Density of monopoles in various gauges}
\begin{figure}[h]
\epsfxsize=15.5cm\epsffile{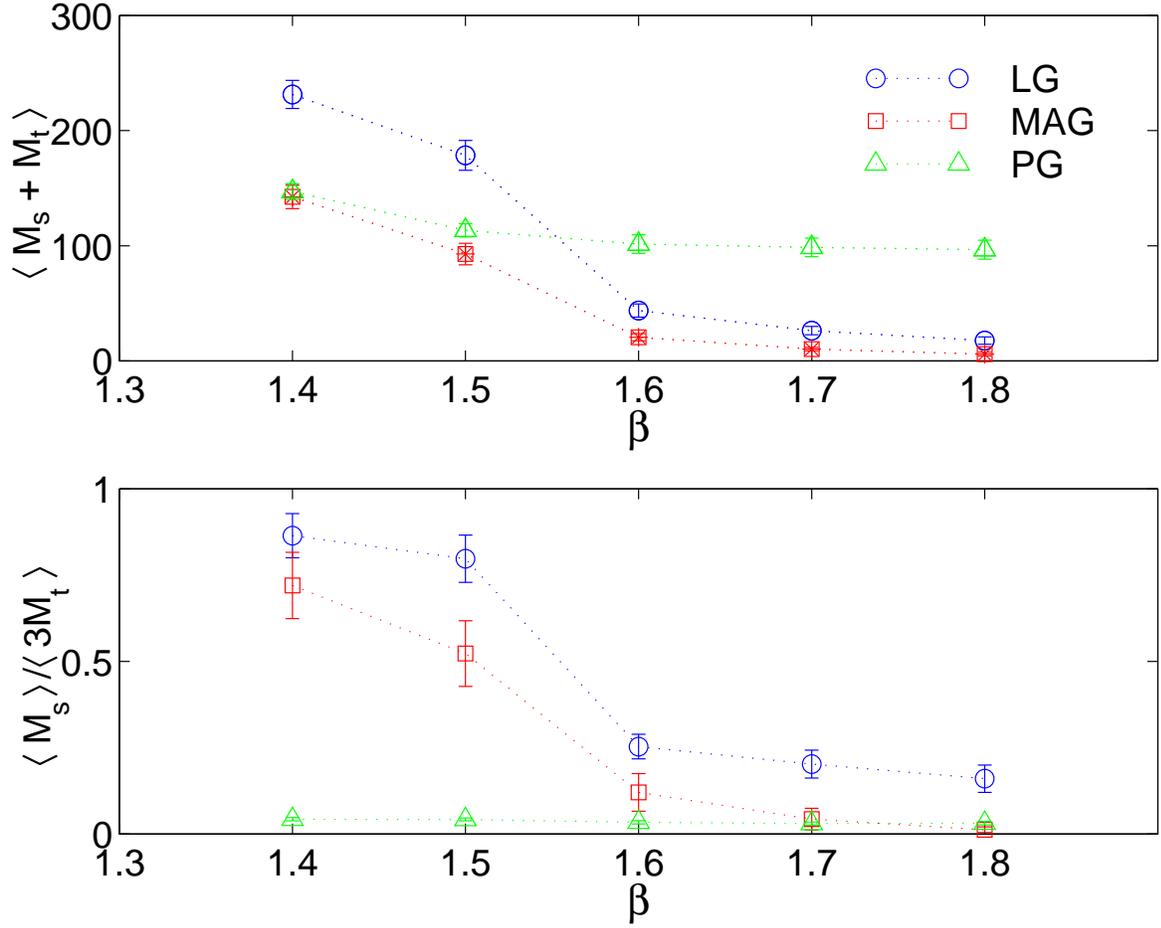}
\caption{ 
      Average total monopole length on the configurations (top) and 
		the space-time asymmetry of magnetic current densities (bottom) 
		shown for three different gauges as a function of $\beta$.
\label{fig:asymmetrie}
}
\end{figure}

In Fig. \ref{fig:asymmetrie} (top)  we show the temperature dependence of the
average number of monopole currents $\langle M_s+M_t \rangle$ 
(total length of all monopole trajectories for a lattice of size 
$12^3 \times 4$), for the different gauges. 
If monopoles are important for confinement, one would 
expect to see a decrease accompanying the phase transition.
In fact, this is only the case for the LG and, less drastically, for the MAG.
For all temperatures the monopole number in LG is bigger than that
in MAG as expected in Ref. \cite{SIJS}.
The monopole number in PG is almost independent of the temperature!

It is interesting to ask for the anisotropy ratio
$\frac{\langle M_s \rangle}{3\langle M_t \rangle} $
as a function of the temperature, shown in Fig. \ref{fig:asymmetrie} (bottom).
At zero temperature this ratio should be equal to one
for isotropic gauges like MAG and LG. But it is
expected to differ from one at higher temperatures.
In our first study of RG smoothing \cite{FEUE} for MAG, we observed
a drop towards the deconfined phase much more pronounced than for 
unsmoothed, Monte Carlo configuration. This is due to the strong isotropic 
noise which is present in the magnetic currents before smoothing.

We find the ratio smaller than one at our different finite
temperatures, and it drops towards and across the deconfining transition.
In the case of MAG, beyond the deconfinement temperature the anisotropy
ratio is approaching zero. For LG the ratio is closer to one in the 
confinement phase and drops more abruptly crossing the transition.
It decreases slowly for $T > T_c$.
We have observed that also the LG monopole network decays
(similar to the MAG monopoles \cite{PRD98}), into a few temporally closed
world lines which are, however, not perfectly static as in the case of MAG.
In view of this, both the MAG and LG data suggest the conclusion that in the
deconfinement phase the corresponding Abelian monopoles become preferably
static objects. What becomes static in the case of the LG are the extended 
``channels'' of small Higgs modulus (see Fig. \ref{fig:Xlokal}).
For MAG, this observation has been first discussed in Ref. \cite{BORN}.

For the PG the anisotropy ratio of the corresponding monopoles does not 
depend on temperature at all. Moreover it is far from isotropy, with
$\frac{\langle M_s \rangle}{3\langle M_t \rangle}\approx0.05$
(almost no spacelike magnetic currents) even in the confinement phase.
Thus, the magnetic currents observed in the PG
are hardly related to the confinement phenomenon.

\begin{figure}[h]
\begin{tabular}{c}
\epsfxsize=15.0cm\epsffile{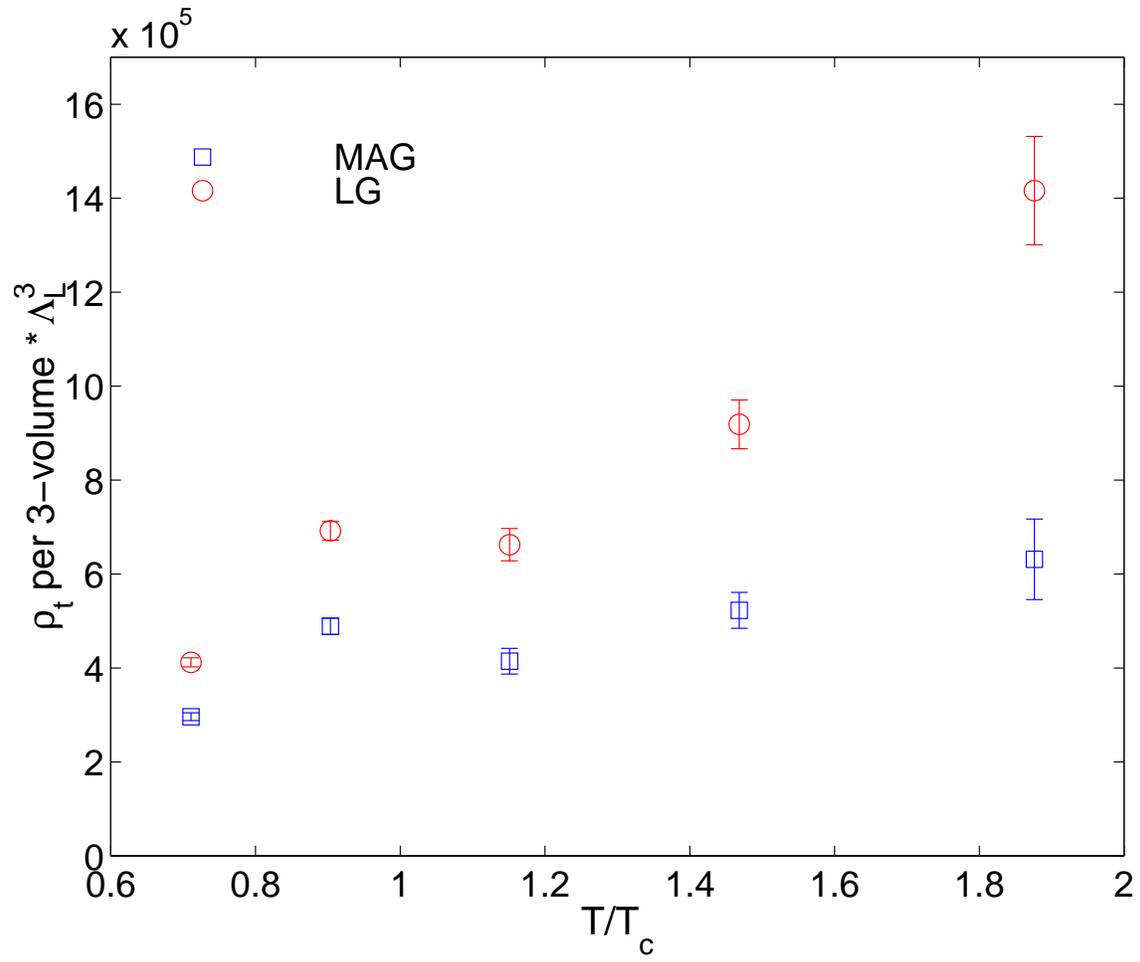}
\end{tabular}
\caption{ 
      Average spatial densities of (timelike) monopole and antimonopole currents
      shown for MAG and LG as a function of physical temperature.  
\label{fig:separate}
}
\end{figure}

For clarity we show in Fig. \ref{fig:separate} the average density
of timelike magnetic currents per $3D$ hyperplane orthogonal to its direction,
as a function of the temperature in units of $\Lambda_L^3$. The upper data
points represent the LG monopoles, the lower the MAG monopoles.
In both of the gauges the physical density of the (almost static)
timelike monopoles starts rising as a function of the temperature 
for $T > 1.2~T_{\mathrm{c}}$. 
It has been speculated that the deconfinement phase is characterized
by a bosonic gas of 't Hooft-Polyakov monopoles with a mass linearly
rising with the temperature \cite{LAURSEN,BORN}.

\subsection{Local properties of monopoles in various gauges}
By definition, DGT monopole currents are localized along the links
of the dual lattice which are related to $3D$ cubes of the original lattice.
They test the properties of the non-Abelian gauge field most locally.
We have defined in Sec. II  local actions and topological charge
densities. Now we want to discuss the different gauges from the point of view
of how their respective monopole trajectories probe the vacuum, 
i.e., whether
these monopoles are equipped with particular, gauge independent properties.
We expect this picture to become clearer when, as in our case, UV
fluctuations are removed from the gauge field configurations.

\begin{figure}[h]
\begin{tabular}{l}
\epsfxsize=12.5cm\epsffile{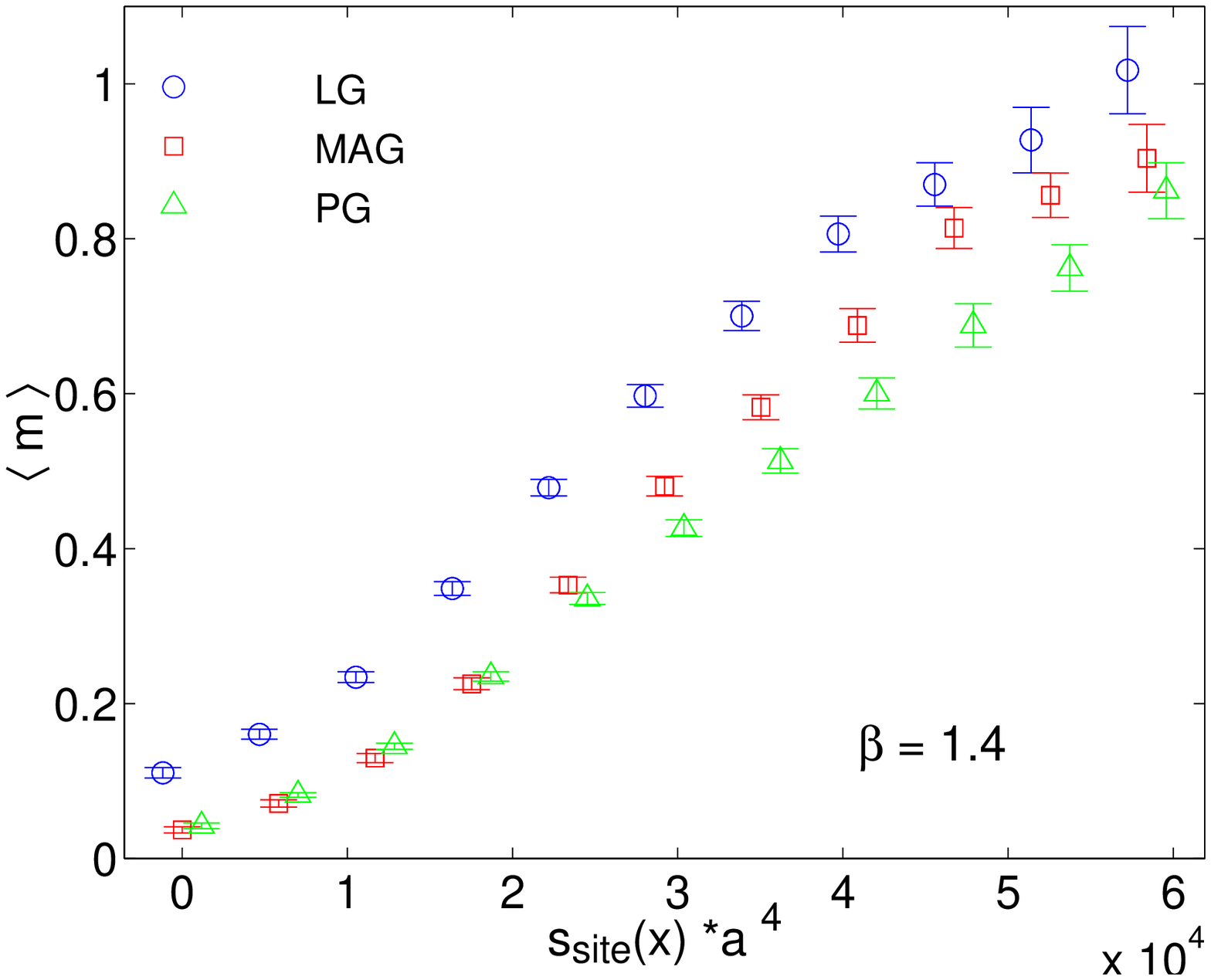}\\
\epsfxsize=11.5cm\epsffile{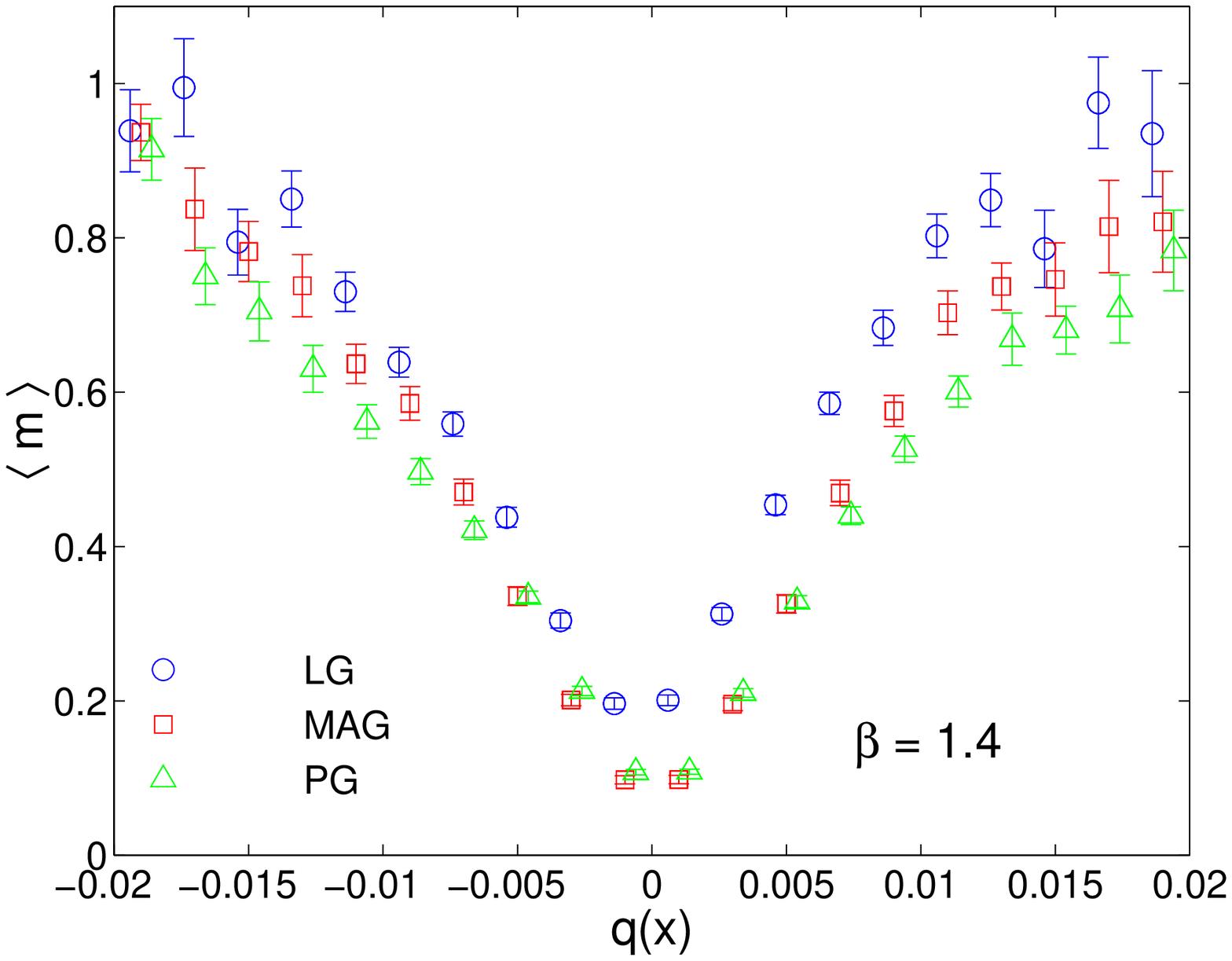}
\end{tabular}
\caption{
      Average number of monopole currents on the nearest dual links for lattice
      points having a value of local action $s_{site}$ within the same 
      respective bin (top). The same for lattice points binned according to 
      topological charge density $q$ (bottom). Both plots refer to $\beta=1.4$ 
      (confinement phase).
\label{fig:actbinmon}
}
\end{figure}

First let us have a look how the average monopole occupation number on the 
dual links near to a site depends
on the local action and charge, in analogy to Fig. \ref{fig:binX}.
Fig. \ref{fig:actbinmon} shows this for the three gauges.
In this respect, no drastic difference between the gauges is seen. 
Again the occupation numbers of LG monopoles are bigger than in the case of
MAG, those for PG somewhat lower. No distinction is made here with respect 
to the direction (spatial vs. temporal) of the magnetic currents.

\begin{figure}[!thb]
\begin{center}
\begin{tabular}{cc}
$\beta=1.4$ &  $\beta=1.8$ \\
\epsfxsize=8.0cm\epsffile{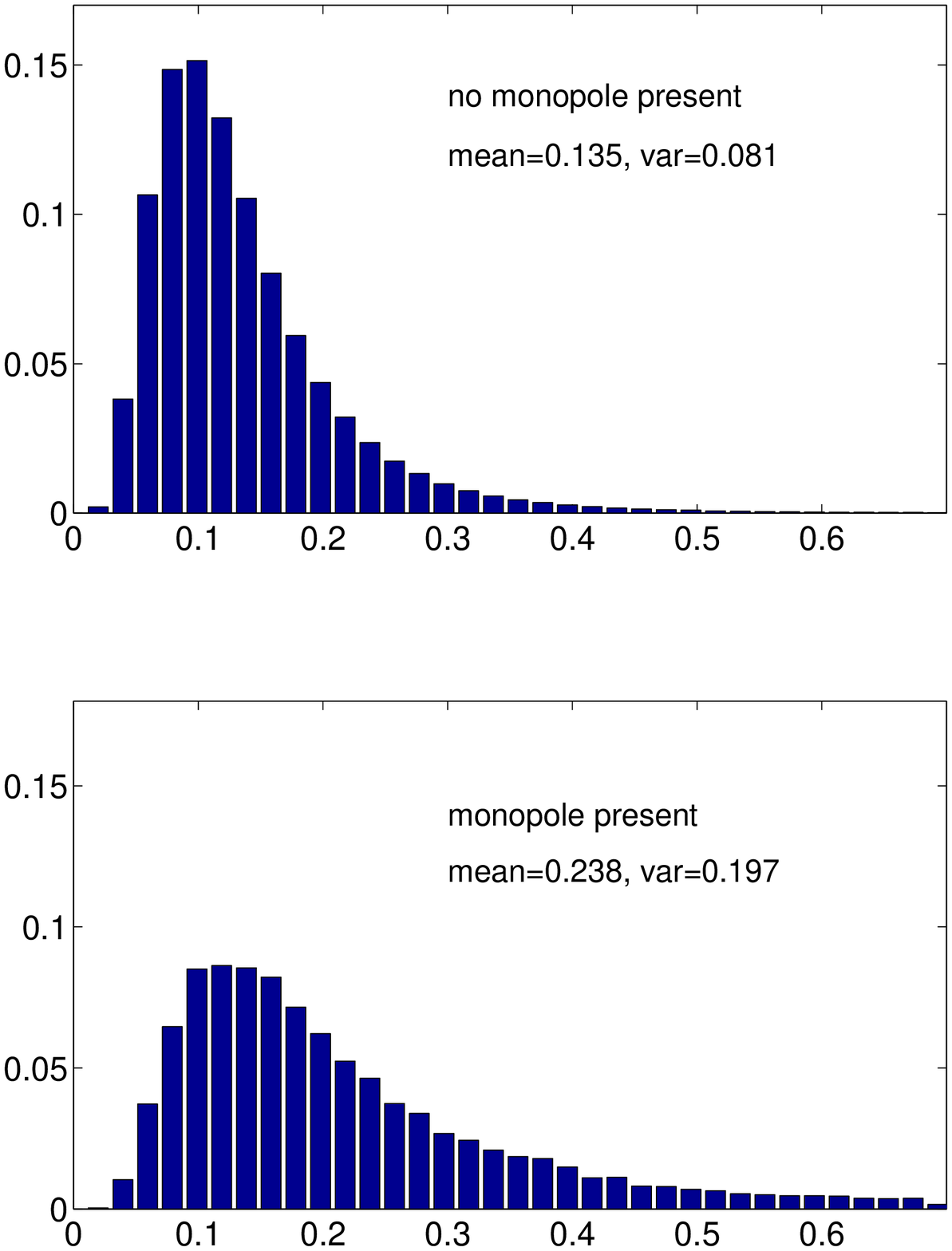} &
\epsfxsize=8.0cm\epsffile{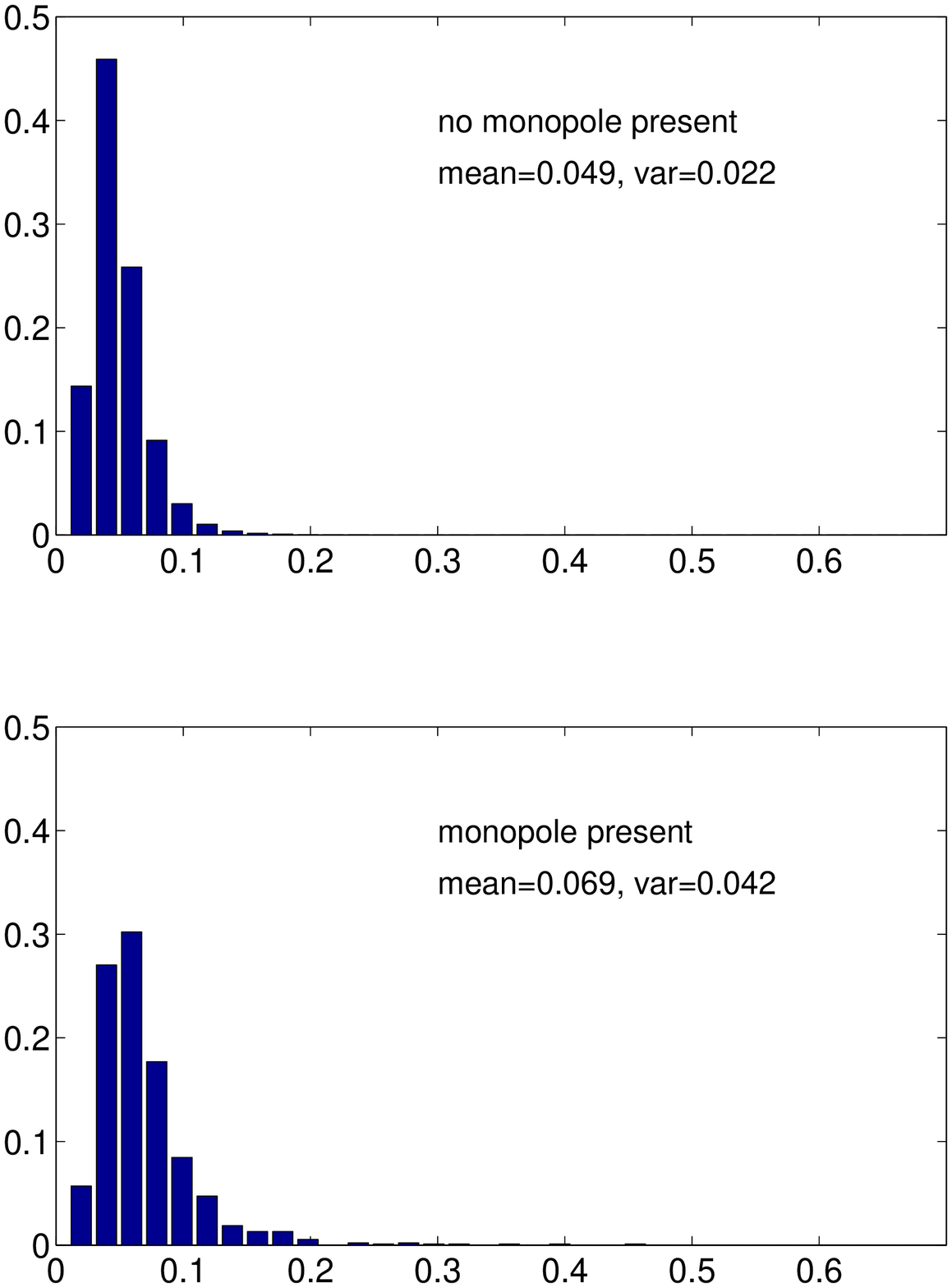} \\
{\small$ S_{3-cube} $}  &{\small $S_{3-cube} $}
\end{tabular}
\end{center}
\caption{
      Probability distribution of cube action $s_{3-cube}$ in the absence (top)
      and in presence (bottom) of a DGT monopole obtained in the LG. 
      $\beta=1.4$ (left) refers to the confinement, $\beta=1.8$ (right) to the 
      deconfinement phase.
\label{fig:probact140}
 }
\end{figure}

In Fig. \ref{fig:probact140} we show the distribution of $s_{3-cube}$, 
separately for cubes free of monopoles (top) and for cubes distinguished 
by monopoles detected in the LG (bottom),
for the lowest and the highest $\beta$-value. This local action contains 
all loops contributing to $S$ localized on the $3D$ cube under consideration.
We see that the distributions for the two cases are different.
Mean and variance in the case of presence of monopoles are almost a
factor of two larger, indicating that on average significantly more action
is picked up per unit of length along an actual monopole trajectory
than would be picked up along a random walk in the given gauge field background.

The situation is very similar for the MAG case, the difference between
the action distributions for monopole-free and monopole-carrying cubes 
is somewhat more pronounced. Actually, these distributions are insensitive
to the {\it direction} of the monopole current only in the confinement phase. 
In the deconfined phase there are important differences. For LG, monopoles 
occupying a cube dual to a spacelike link (which are still frequent in the 
deconfined phase, in contrast to MAG!)
are accompanied by a probability distribution of local action which is
similar to that of an empty cube, and for timelike ones the action is not 
as high as in the MAG case. This explains the higher multiplicity of LG 
monopoles and the lack of suppression of spacelike ``detours''
for world lines closing in timelike direction.
For cubes along the trajectories of PG monopoles 
the action distribution is practically the same as for the
empty cubes. This holds in both phases.
Looking back to section III.C, we conclude that among the  
gauge invariant quantities $X^2$ is the observable, whose distribution 
reflects the absence or presence of monopoles most clearly. This holds 
independently of the direction of the monopole current in both phases
(Fig. \ref{fig:probX}).

\subsection{Excess action and charge}
\begin{figure}
\begin{tabular}{ c }
\epsfxsize=12.0cm\epsffile{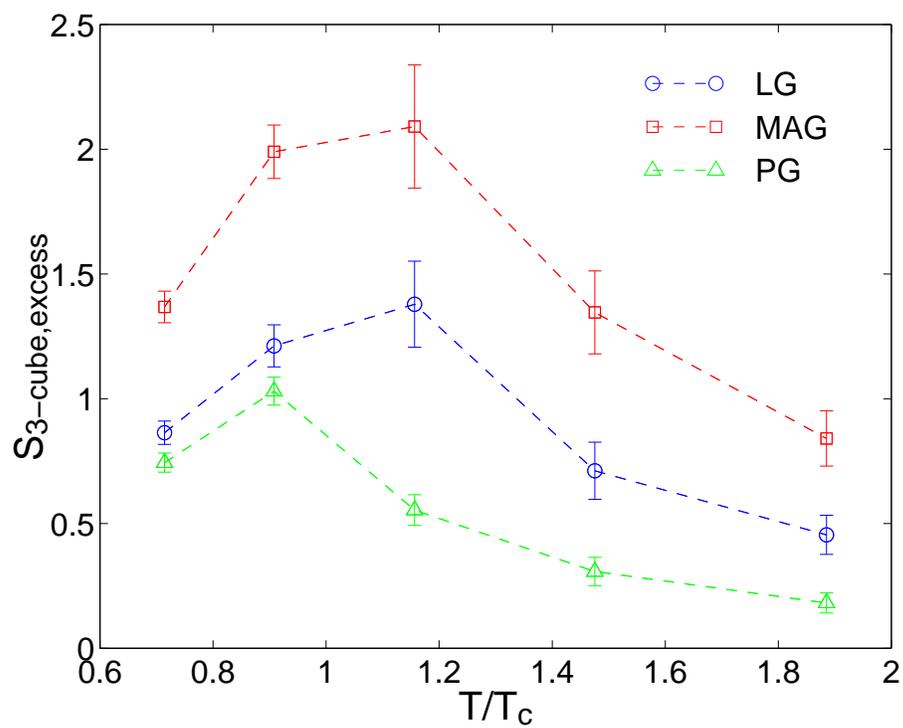} \\
\epsfxsize=12.0cm\epsffile{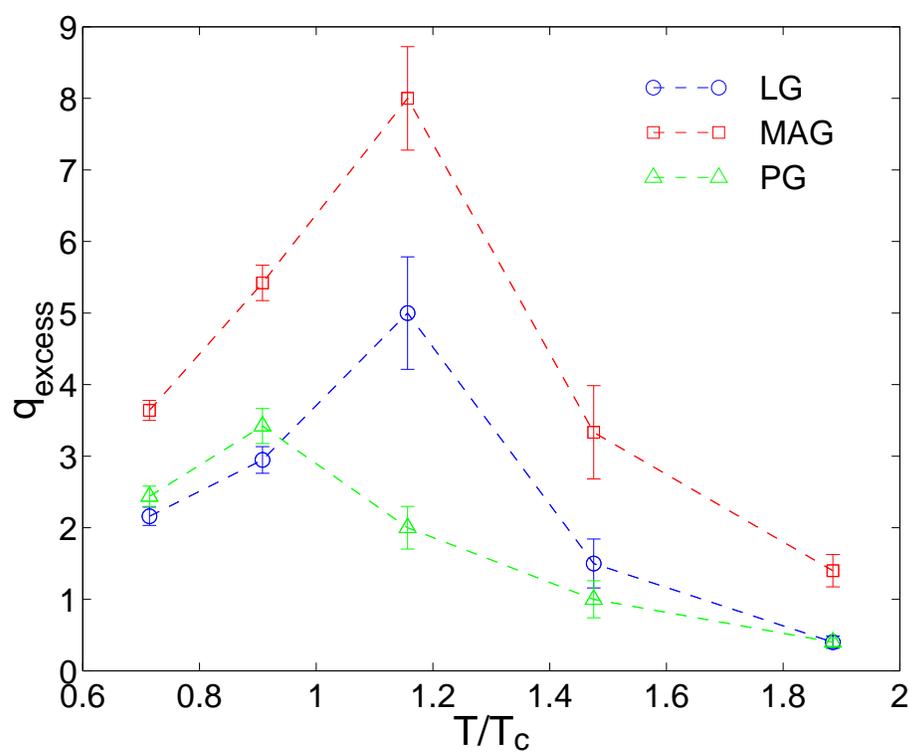}
\end{tabular}
\caption{
      Excess action (top) and topological charge (bottom) shown as function of
      physical temperature.
\label{fig:excess}
}
\end{figure}

In a more concise way than by the distributions of local action just discussed
we can characterize the different types of monopoles and the change with
$\beta$ by an average excess action of monopoles defined as 
\begin{equation}
S_{\mathrm{ex}}= \frac{< S_{\mathrm{monopole}}-S_{\mathrm{no~monopole}} > }
       {<S_{\mathrm{no~monopole}} >}  \,\, ,
\end{equation}
where $S_{\mathrm{monopole}}$ is again the action localized on a
three-dimensional cube mentioned before.
Replacing the action in the above expression by the modulus of the
topological charge density according to the L\"uscher method
we can define the charge excess $q_{\mathrm{ex}}$ compared with
the noise of $|q(x)|$ anywhere else in the vacuum.
For details of the definition of the local operators see Ref. \cite{PRD98}.
Fig. \ref{fig:excess} shows that already just below $T_{\mathrm{c}}$
the excess action for MAG and LG monopoles is clearly above one
(indicating an excess of action of more than a factor of two compared to the
bulk average) and rises across the transition. Around $T_{\mathrm{c}}$,
the excess charge is even a few times as big for MAG and LG monopoles.
These results are more pronounced compared with a $T=0$ study 
\cite{BAKK98} (using Wilson action, without cooling or smoothing)
and emphasize the dynamical importance of these two types of monopoles
at the deconfining transition.

\section{Does Abelian dominance of the string tension hold in all gauges?}
The Abelian dominance of the string tension \cite{SuzYot} in the case
of MAG was a strong argument for choosing this gauge to look deeper
for the role of monopoles. Finally, this has given rise
to the construction of infrared effective actions for QCD \cite{SuzAct}.
Originally, of course, the concept of Abelian dominance 
was referring to generic Monte Carlo configurations. As a by-product
of our study on monopoles in various gauges, we present here results 
concerning this question for smoothed configurations. 

\begin{figure}
\begin{center}
\epsfxsize=15.0cm\epsffile{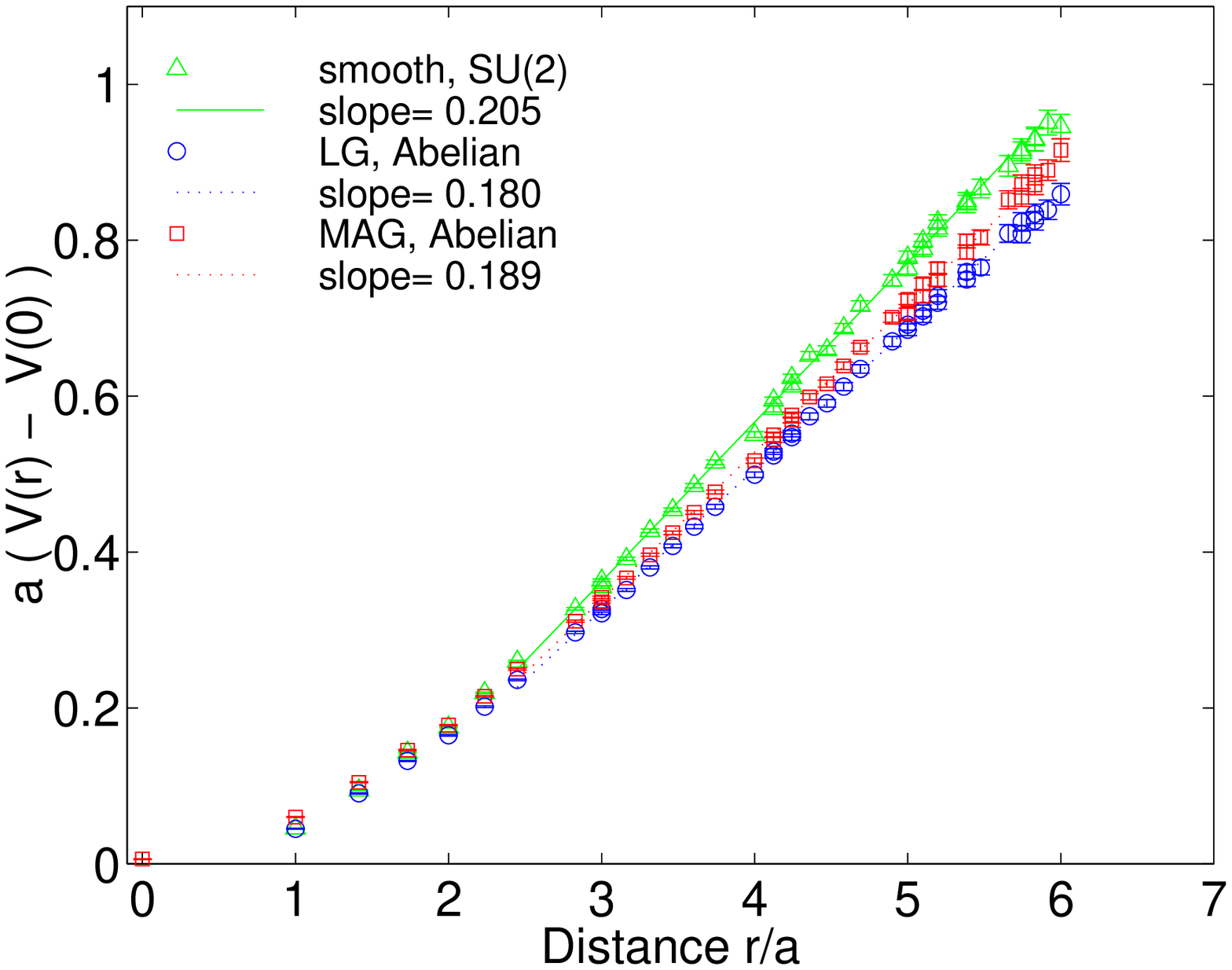}
\end{center}
\caption{ 
      Heavy quark potential evaluated after smoothing from non-Abelian 
      $SU(2)$ Polyakov line correlators at $\beta =1.4$ (confinement) and from 
      the Abelian Polyakov line correlators in Abelian projections from 
      MAG and LG, respectively. For PG there is no difference with the 
      non-Abelian potential.
\label{fig:LLpot}
}
\end{figure}

RG smoothing preserves the non-Abelian string tension 
within $7$ \% and removes the Coulombic part of the heavy quark force
at small distance.
In Fig. \ref{fig:LLpot} we display the static quark-antiquark potential 
as obtained from Polyakov line correlators, for the smoothed $SU(2)$ fields
and, in the case of LG and MAG, after Abelian projection. The corresponding 
values for the string tensions are found in Table \ref{tension}.
The Abelian string tension of the MAG is about $5$ \% less than that
calculated for the non-Abelian $SU(2)$ field. The Abelian string tension of 
LG is again somewhat smaller than for the MAG. In qualitative respect, 
we still find Abelian dominance for smoothed lattice fields, independent
of the gauge chosen for projection. Notice, however, that in the PG 
(where the temporal links $U_{({\bf x},x_4),4}$ 
become diagonal and $x_4$ independent)
the non-Abelian string tension extracted from Polyakov line correlators
is trivially identical with the Abelian one.

\section{Conclusion}

We have collected additional evidence that the RG smoothing technique 
(with an approximate classically perfect action) 
is a powerful tool, here used to investigate semiclassical aspects 
of the monopole structure of the Yang-Mills vacuum.
We have presented further results concerning the recently proposed, 
physically better motivated and gauge invariant method to
identify monopoles on the lattice, the so-called Laplacian gauge (LG). 
We contrasted the corresponding monopoles with other types of monopole 
trajectories which usually are obtained in a gauge dependent way.
What parallels exist and how different the monopole content can be if
different gauges are used to localize them
has been examplified by considering three gauges (MAG, LG and PG).
In a first step we showed that the gauge invariant way to localize monopoles 
is consistent with the DeGrand-Toussaint (DGT) 
method being applied to Abelian projected configurations after 
LG fixing. This correspondence becomes 
clearer if smoothed Monte Carlo configurations are analyzed.
This justifies to take over the DGT method to study the 
monopole content of configurations also in the LG, at least after RG smoothing.
Accepting this technique, we also found strong correlations between
positions of MAG and LG monopoles.  

While MAG and LG monopoles (as extracted by the DGT method) behave similar 
at the deconfining phase transition concerning density and anisotropy,
monopoles identified in the Polyakov gauge apparently lack a dynamical 
relevance for confinement and at the deconfinement transition.
In a second step we then analyzed
trajectories of monopoles identified in various gauges and found that 
monopoles (in all three gauges studied) appear preferably in regions which 
are characterized by enhanced action and topological charge density. 
The reverse is not universally true:
only monopoles detected in LG and MAG are characterized
by significantly different local distributions of action.
We have shown earlier \cite{PRD98} that
strong gauge fields are locally (anti)selfdual.
This is further, circumstantial evidence that the monopoles are in fact dyons
in the confinement phase. As a by-product of our study we found 
that almost the complete string tension
can be recovered from the Abelian projected field corresponding to
various Abelian gauges, including LG and PG. For the latter
this is trivially true, as long as we measure the string tension by means
of the Polyakov line correlator.

We have shown that for smoothed gauge field background
monopole trajectories carry an excess action of about twice the bulk
average of local action.
Similarly, monopoles also carry an even bigger excess topological charge.
In the confinement phase this observation  
applies to monopoles independent of which
gauge has been chosen for identifying them,
but PG monopoles are very different in the deconfinement phase.
We conclude that the DGT monopoles related to MAG and LG 
behave similar physically and can be semiclassically 
interpreted as physical objects which carry considerable 
action and topological charge in the vicinity of the deconfinement
transition.

This work was supported in part by FWF under project P11456.
One of us (M. M.-P.) acknowledges support by the European TMR network 
{\it Phase Transitions in Hot Matter} under contract FMRX-CT97-0122.
E.-M. I. wishes to thank T. Suzuki and M. I. Polikarpov for discussions
on dual descriptions of confinement.


\newpage
\begin{table}[h]
\begin{center}
\begin{tabular}{ l c c c c }
$w(i,j)$            & $j=1$   &  $j=2$  & $j=3$ & $j=4$   \\
\hline
$i=1$ (plaquettes)  & $1.115504$& $-.5424815$ & $.1845878$ & $-.01197482$ \\
$i=2$ (6-link loops)& $-.01443798$& $.1386238 $ & $-.07551325 $ & $.01579434$ 
\\
\end{tabular}
\end{center}
\caption{ Weight coefficients of the simplified fixed-point action.}
\label{tab:weights}
\end{table}

\newpage

\begin{table}
\begin{tabular}{ll}
      variants of pure $SU(2)$ theory            &  $\sigma~a^2$ \\
\hline
      non-Abelian, without smoothing             &  0.22(2)      \\
      non-Abelian, with RG smoothing             &  0.205(2)     \\
      RG smoothed, projected to Abelian from MAG &  0.189(1)     \\
      RG smoothed, projected to Abelian from LG  &  0.180(2)     \\
      RG smoothed, projected to Abelian from PG  &  0.205(2)     \\
\end{tabular}
\caption{
      \label{tension}
      String tensions extracted from Polyakov line correlators
      at $\beta=1.4$.
}
\end{table}


\newpage

\newpage

\newpage

\newpage

\newpage

\newpage

\newpage

\newpage

\newpage

\newpage

\newpage

\newpage

\begin{thebibliography}{9}


\bibitem{ILM}      E. Shuryak,
                   \Journal{\NPB}{B203}{93}{1982};
                   T. Sch\"afer and E. Shuryak,
                   \Journal{Rev. Mod. Phys.}{70}{323}{1998}.
\bibitem{DSC1}      G. 't Hooft,
                   \Journal{\NPB}{B190}{455}{1981}.
\bibitem{DSC2}     S. Mandelstam,
                   \Journal{Phys. Rep.}{C23}{245}{1976}.
\bibitem{NEGELE98} J. Negele, talk presented at {\em Lattice 98}.
\bibitem{ADHM}     S. Maedan and T. Suzuki,
		   \Journal{Prog. Theor. Phys.}{81}{229}{1989}; 
		   \Journal{Prog. Theor. Phys.}{84}{130}{1990};
                   S. Kamizawa, Y. Matsubara, H. Shiba and T. Suzuki,
                   \Journal{\NPB}{B389}{563}{1993}. 
\bibitem{SmitVdS}  J. Smit and A. J. van der Sijs, 
                   \Journal{\NPB}{B355}{603}{1991}.
\bibitem{SuzYot}   T. Suzuki and I. Yotsuyanagi,
                   \Journal{\PRD}{42}{4257}{1990};
		   S. Hioki, S. Kitahara, S. Kiura, Y. Matsubara, 
		   O. Miyamura, S. Ohno and T. Suzuki,
                   \Journal{\PLB}{B272}{326}{1991}.
\bibitem{BORN}     V. G. Bornyakov, V. K. Mitrjushkin 
		   and M. M\"uller-Preussker,
                   \Journal{\PLB}{B284}{99}{1992}.
\bibitem{ShiSuz}   H. Shiba and T. Suzuki,
                   \Journal{\PLB}{B333}{461}{1994}.
\bibitem{BALI}     G. Bali, V. G. Bornyakov, M. M\"uller-Preussker and 
                   K. Schilling,
                   \Journal{\PRD}{54}{2863}{1996}.
\bibitem{SchiBo}   V. G. Bornyakov and G. Schierholz,
	           \Journal{\PLB}{B384}{190}{1996}. 
\bibitem{MIYAMURA} O. Miyamura,
                   \Journal{\PLB}{B353}{91}{1995}; 
                   \Journal{Prog. Theor. Phys. Suppl.}{120}{171}{1995}. 
\bibitem{SasaMiya} S. Sasaki and O. Miyamura,
                   \Journal{Nucl. Phys. {\bf B} (Proc. Suppl.)}{63}{507}{1998};
                   \Journal{\PLB}{B443}{331}{1998}; 
                   \Journal{\PRD}{59}{094507}{1999}. 
\bibitem{SuzAct}   H. Shiba and T. Suzuki,
                   \Journal{\PLB}{B351}{519}{1995}; 
                   S. Kato, N. Nakamura, T. Suzuki and S. Kitahara,
                   \Journal{\NPB}{B520}{323}{1998}; 
                   M. N. Chernodub, S. Kato, N. Nakamura, M. I. Polikarpov
		   and T. Suzuki, e-Print Archive: hep-lat/9902013. 
\bibitem{THUR95}   S. Thurner, H. Markum and W. Sakuler,
                   in {\em Proceedings of Confinement 95}
                   (World Scientific, Singapore, 1995).
\bibitem{THUR96}   S. Thurner, M. Feurstein, H. Markum and W. Sakuler,
                   \Journal{\PRD}{54}{3457}{1996}.
\bibitem{SUGA97}   H. Suganuma, S. Sasaki, H. Ichie, F. Araki and O. Miyamura,
                   \Journal{Nucl. Phys. {\bf B} (Proc. Suppl.)}{53}{528}{1997}.
\bibitem{CHER95}   M. N. Chernodub and F. V. Gubarev,
                   \Journal{JETP Lett.}{62}{100}{1995}.
\bibitem{BROW97}   R. C. Brower, K. N. Orginos and C.-I. Tan,
                   \Journal{\PRD}{55}{6313}{1997}.
\bibitem{REIN97}   H. Reinhardt,
                   \Journal{\NPB}{B503}{505}{1997}.
\bibitem{WIPF}     C. Ford, U. G. Mitreuter, T. Tok, A. Wipf 
		   and J. M. Pawlowski,
                   \Journal{Ann. of Phys.}{269}{26}{1999}; 
                   C. Ford, T. Tok and A. Wipf,
                   e-Print Archive: hep-th/9809209, hep-th/9811248. 
\bibitem{LENZ}     O. Jahn and F. Lenz,  
                   \Journal{\PRD}{58}{085006}{1998}. 
\bibitem{KRON87}   A. S. Kronfeld, G. Schierholz and U.-J. Wiese,
                   \Journal{\NPB}{B293}{461}{1987}.
\bibitem{degrandT} T. A. DeGrand and D. Toussaint,
                   \Journal{\PRD}{22}{2478}{1980}.
\bibitem{VINK}     J. C. Vink and U.-J. Wiese,
                   \Journal{\PLB}{B289}{122}{1992}. 
\bibitem{SIJS}     A. van der Sijs,
                   \Journal{Nucl. Phys. {\bf B} (Proc. Suppl.)}{53}{535}{1997};
                   \Journal{Prog. Theor. Phys. Suppl.}{131}{149}{1998}.
\bibitem{FEUE}     M. Feurstein, E.-M. Ilgenfritz,
                   M. M\"uller-Preussker and  S. Thurner,
                   \Journal{\NPB}{B511}{421}{1998}.
\bibitem{fermionic} R. G. Edwards, U. M. Heller and R. Narayanan,
                   \Journal{\NPB}{B535}{403}{1998}.
\bibitem{degrand}  T. A. DeGrand, A. Hasenfratz and De-cai~Zhu,
                   \Journal{\NPB}{B475}{321}{1996};
                   \Journal{\NPB}{B478}{349}{1996}.
\bibitem{PRD98}    E.-M. Ilgenfritz, H. Markum,
                   M. M\"uller-Preussker and S. Thurner,
                   \Journal{\PRD}{58}{094502}{1998}.
\bibitem{coolimp}  P. de Forcrand, M. Garcia Perez and I.-O. Stamatescu, 
                   \Journal{\NPB}{B499}{409}{1997}.
\bibitem{HASE}     P. Hasenfratz and F. Niedermayer,
                   \Journal{\NPB}{B414}{785}{1994};
                   T. DeGrand, A. Hasenfratz, P. Hasenfratz and F. Niedermayer,
                   \Journal{\NPB}{B454}{578}{1995}; 
		   \Journal{\NPB}{B454}{615}{1995}.
\bibitem{boulder_cycling} T. DeGrand, A. Hasenfratz and T. G. Kovacs,
                   \Journal{\NPB}{B505}{417}{1997}.
\bibitem{luescher} M. L\"uscher, 
	           \Journal{Comm. Math. Phys.}{85}{29}{1982};
                   I. A. Fox, J. P. Gilchrist, M. L. Laursen and G. Schierholz,
                   \Journal{\PRL}{54}{749}{1985}.
\bibitem{letter96} M. Feurstein, H. Markum and S. Thurner,
	           \Journal{\PLB}{B396}{203}{1997}. 
\bibitem{Seixas}   G. Schierholz, J. Seixas and M. Teper,
	           \Journal{\PLB}{B157}{209}{1985}.
\bibitem{HOLLANDS} S. Hollands and M. M\"uller-Preussker,
	           e-Print Archive: hep-th/9901114.
\bibitem{LAURSEN}  M. L. Laursen and G. Schierholz,
                   \Journal{\ZPC}{38}{501}{1988}.
\bibitem{BAKK98}   B. L. G. Bakker, M. N. Chernodub and M. I. Polikarpov,
                   \Journal{\PRL}{80}{30}{1998}.

\end{thebibliography}
\end{document}